\documentclass{article}
\usepackage{spconf,amsmath,graphicx}

\usepackage{multirow}
\usepackage{graphicx}
\usepackage{amssymb, amsmath, bm, mathtools,array}
\usepackage{textcomp}
\usepackage{hyperref}
\usepackage{verbatim,lipsum}
\usepackage{booktabs}
\usepackage{tcolorbox}
\usepackage{lscape}
\usepackage{subcaption}
\usepackage{multirow}
\usepackage{colortbl}
\usepackage{color}
\usepackage{multibib}

\RequirePackage{color}\definecolor{BLUE}{rgb}{0,0,0}

\newcommand{\mA}{{\texttt{LFCC-LCNN}}}
\newcommand{\mB}{{\texttt{AASIST}}}
\newcommand{\mC}{{\texttt{W2V-GF}}}

\newcommand{\setTLA}{{\shortstack{\texttt{LA19} \\ \texttt{trn}}}}
\newcommand{\setTWF}{{\shortstack{\texttt{WF} \\ \texttt{trn}}}}
\newcommand{\setTVI}{{\shortstack{\texttt{Voc.} \\  \texttt{v1}}}}

\newcommand{\setTVIII}{{\shortstack{\texttt{Voc.} \\ \texttt{v2}}}}
\newcommand{\setTVIV}{{\shortstack{\texttt{Voc.} \\ \texttt{v3}}}}
\newcommand{\setTVV}{{\shortstack{\texttt{Voc.} \\ \texttt{v4}}}}

\newcommand{\setTLAshort}{{\texttt{LA19trn}}}
\newcommand{\setTWFshort}{{\texttt{WFtrn}}}
\newcommand{\setTVIshort}{{\texttt{Voc.v1}}}
\newcommand{\setTVIIshort}{{\texttt{Voc.v2}}}
\newcommand{\setTVIIIshort}{{\texttt{Voc.v2}}}
\newcommand{\setTVIVshort}{{\texttt{Voc.v3}}}
\newcommand{\setTVVshort}{{\texttt{Voc.v4}}}
\newcommand{\setTVVIIshort}{{\texttt{Voc.v5}}}

\newcommand{\setELAold}{{\texttt{LA15eval}}}
\newcommand{\setELA}{{\texttt{LA19eval}}}
\newcommand{\setELAII}{{\texttt{LA21eval}}}
\newcommand{\setEDF}{{\texttt{DF21eval}}}
\newcommand{\setELATRIM}{{\texttt{LA19etrim}}}
\newcommand{\setELAHID}{{\texttt{LA21hid}}}
\newcommand{\setEDFHID}{{\texttt{DF21hid}}}
\newcommand{\setEWFE}{{\texttt{WaveFake}}}
\newcommand{\setEDFE}{{\texttt{InWild}}}

\newcommand{\setGAll}{{\texttt{Pooled}}}

\newcommand{\selfcircle}[1]{\raisebox{.5pt}{\textcircled{\raisebox{-.9pt} {#1}}}}

\newcites{app}{Appendix References}

\title{Spoofed training data for speech spoofing countermeasure can be efficiently created using neural vocoders}
\name{Xin Wang\thanks{This study is supported by JST CREST Grants JPMJCR18A6 and JPMJCR20D3, MEXT KAKENHI Grants 21K17775 and 21H04906, and Google AI for Japan program.}, Junichi Yamagishi}
\address{National Institute of Informatics, Japan }

\begin{document}
\ninept

%%%%% Copyright notice BEGIN
\onecolumn
{\noindent\Large \textbf{IEEE Copyright Notice}}

${}$

{\noindent\large \copyright 2023 IEEE. 
Personal use of this material is permitted. Permission from IEEE must be obtained for all other uses, in any current or future media, including reprinting/republishing this material for advertising or promotional purposes, creating new collective works, for resale or redistribution to servers or lists, or reuse of any copyrighted component of this work in other works.

${}$

\noindent
Accepted by 2023 IEEE International Conference on Acoustics, Speech and Signal Processing

${}$

\noindent
DOI: to appear
}
%%%%% Copyright notice END

\twocolumn
\maketitle
\begin{abstract}
A good training set for speech spoofing countermeasures requires diverse TTS and VC spoofing attacks, but 
generating TTS and VC spoofed trials for a target speaker may be technically demanding. 
Instead of using full-fledged TTS and VC systems, this study uses neural-network-based vocoders to do copy-synthesis on bona fide utterances. The output data can be used as spoofed data. To make better use of pairs of bona fide and spoofed data, this study introduces a contrastive feature loss that can be plugged into the standard training criterion. 
On the basis of the bona fide trials from the ASVspoof 2019 logical access training set, this study empirically compared a few training sets created in the proposed manner using a few neural non-autoregressive vocoders. 
Results on multiple test sets suggest good practices such as fine-tuning neural vocoders using bona fide data from the target domain. The results also demonstrated the effectiveness of the contrastive feature loss. Combining the best practices, the trained CM achieved overall competitive performance. Its EERs on the ASVspoof 2021 hidden subsets also outperformed the top-1 challenge submission.

\end{abstract}
\section{Introduction}
\label{sec:intro}

The detection of speech produced by text-to-speech (TTS) and voice conversion (VC) systems is a well-established but unresolved research topic \cite{evans2013spoofing, LiuASVspoof2021}. 
Most studies use machine-learning-based approaches, and they train a classification model  (a.k.a spoofing countermeasure (CM)) on a training set that contains both human (bona fide) and synthesized (spoofed) speech data. The trained CM is then used to detect unseen test data.  

Creating high-quality and diverse spoofing data requires tremendous effort. For example, the ASVspoof 2019 challenge logical access (LA) database \cite{asvspoof2019_database} was constructed with the help of more than ten organizations over six months. Furthermore, spoofing methods are constantly evolving. 
For real applications, practitioners may need to update the CM training data from time to time.
Although some TTS and VC algorithms with zero-shot learning  \cite{cooper2020zero, casanova2022yourtts} make it easier to generate speech for any speaker, they cover only part of the existing TTS-VC paradigms. Collecting spoofed data from diverse TTS and VC systems is still demanding.

This study investigates an alternative way to create spoofed data for CM training. Instead of using full-fledged TTS and VC systems, this study uses the \emph{vocoder}, a last-step module in both TTS and VC systems, to create spoofed data in a copy-synthesized manner.  Specifically, acoustic features such as Mel-spectrogram are extracted from a bona fide waveform and fed to a vocoder to re-synthesize the waveform. The copy-synthesized or \emph{vocoded} waveform is treated as spoofed data. While vocoded data is different from TTS/VC-synthesized data because the latter is generated given predicted acoustic features, both carry artifacts inherent to the vocoder. Thus, it is presumably possible to train a CM using vocoded spoofed data and use it to detect spoofed data from actual TTS/VC systems. 
Furthermore, vocoded data is easier to obtain than TTS/VC-synthesized data since most vocoders do not require text transcription, grapheme-to-phoneme conversion, or speaker embeddings. Training a universal vocoder is also comparably easier than training a TTS/VC system.

Vocoded spoofed data has been used for CM training in the past. In \cite{wu2013synthetic, sanchez2014cross, sizov2015joint}, vocoded data produced by digital-signal-processing (DSP)-based vocoders was used to teach a CM to spot phase distortion caused by these vocoders. A more recent database called WaveFake gathers vocoded spoofed data from advanced deep-neural-network (DNN)-based vocoders \cite{frank2021wavefake}, but the data is only from two speakers. Neither is it reported how a CM trained on WaveFake generalizes to other test sets. 

In this study, we plan to answer three questions:
\textbf{1)} Is there any caveat when using the latest neural vocoders to create useful vocoded spoofed data? \textbf{2)} Is there any method that can make good use of bona fide and vocoded data pairs and better train the CM? \textbf{3)} How well does the CM  generalize to actual and advanced TTS and VC spoofing attacks?  
On the basis of the bona fide data in the ASVspoof 2019 training set \cite{asvspoof2019_database}, we build a few training sets with vocoded spoofed data to address the first question. We then introduce a contrastive feature loss to address the second question.
Answers to all the questions are examined through experiments that used the prepared training sets and multiple benchmark test sets with diverse TTS and VC attacks.  The results suggest a few practices for training set creation. They also demonstrate that a CM trained using the best training set in this study and the contrastive feature loss performed well on detecting challenging spoofing attacks and in a few cases better than other CMs in the literature.

\section{Methods}
\label{sec:method}

\subsection{Creating Spoofed Data Using Vocoders}
\label{sec:database}

Creating vocoded data is straightforward: collecting natural (bona fide) speech data, extracting acoustic features, and driving the vocoder to synthesize waveforms. Many open-sourced and pre-trained vocoders can be used off-the-self, fine-tuned or re-trained from scratch.  
However, there are factors we should be aware of. For example, which vocoder should be included? Is there any mismatch between the sampling rate of the bona fide data and that required by the vocoder? Is there any domain mismatch between the vocoder's training data and the bona fide data for copy-synthesis? 

The first factor is also a practical one because the generation speed of some DNN-based vocoders (e.g., autoregressive (AR) models like WaveNet \cite{oord2016wavenet}) is more than $100$ times slower than real-time \cite{yamamoto2020parallel}. Most users, including us, cannot afford the time to generate even a small amount of vocoded data.  Therefore, we investigate only non-AR neural vocoders:
\begin{itemize} 
\item General-adversarial-network-based: HiFiGAN \cite{NEURIPS2020_c5d73680}, Parallel WaveGAN (PWG) \cite{yamamoto2020parallel}, MultiBand MelGAN \cite{yang2021multi};
\item Fusion of DNN and DSP: Harmonic-plus-noise neural source-filter model (Hn-NSF) \cite{wangNSFall}\footnote{This Hn-NSF is different from the NSF vocoder in spoofing attack A08 of the ASVspoof 2019 LA test set. The former uses simplified neural filters, while the latter uses a WaveNet-like network structure (i.e., b-NSF in \cite{wangNSFall}).}, combination of Hn-NSF and HiFiGAN (NSF-HiFiGAN) \cite{tomashenko2022voiceprivacy};
\item Flow-based: WaveGlow \cite{prenger2018waveglow}.
\end{itemize}
Another reason to choose the above vocoders is that they have reliable implementations released. 

As for other factors such as sampling rate and data domain mismatch, we created a few training sets with vocoded spoofed data and empirically investigated what caveat should be cautioned against. Specifically, we used the bona fide data from the ASVspoof 2019 LA training set ({\setTLAshort}) and prepared different versions of vocoded spoofed data using the vocoders above. The created training sets are listed in Table~\ref{tab:tablesets} and described below:
\begin{itemize}
\item {\setTVIshort} follows the WaveFake training subset ({\setTWFshort})\footnote{WaveFake has no official training subset. Hence, we extracted one with around 4,000 randomly selected bona fide English trials and their vocoded data from four vocoders. Details are listed in Table~\ref{tab:training_data}.} and used four vocoders pre-trained by ESPNet \cite{hayashi2020espnet} on a multi-speaker database called LibriTTS \cite{zen2019libritts}. However, since these vocoders operate at a sampling rate of 24 kHz,  the 16-kHz bona fide data to be vocoded was up-sampled to 24 kHz, and the vocoded data was down-sampled back to 16 kHz. 
\item {\setTVIIshort} requires no change of sampling rate during copy-synthesis. The spoofed data was created using our own implemented vocoders trained on LibriTTS and operating at 16 kHz. The vocoders cover the three categories listed above.
\item {\setTVIVshort} is similar to {\setTVIIshort}, but vocoders were trained from scratch on the bona fide data of {\setTLAshort}.
\item {\setTVVshort} is similar to {\setTVIIshort}, but vocoders from {\setTVIIshort} were further trained on the bona fide data of {\setTLAshort}.
\end{itemize} 
Note that all the above training sets have the same amount of bona fide trials as {\setTLAshort}.
Comparisons among {\setTVIIshort},  {\setTVIVshort}, and {\setTVVshort} are expected to show the impact of data domain mismatch, while comparison between {\setTVIIshort} and {\setTVIshort} can be helpful for understanding the impact of the sampling rate.

\subsection{Contrastive Feature Loss for  Bona Fide and Vocoded Data}
\label{sec:sup}

Vocoded spoofed data has (almost) the same duration and tempo as the corresponding bona fide utterance, which has at least two merits. 
First, potential data artifacts caused by an imbalanced distribution of duration in bona fide and spoofed data (e.g., length of non-speech regions \cite{LiuASVspoof2021, muller21_asvspoof}) can be avoided. 
Second, acoustic feature sequences extracted from spoofed and bona fide utterances are aligned with each other, and useful contrastive information can be derived. 

To make good use of the contrastive information, we introduce an auxiliary contrastive feature loss for CM training.
Suppose that we have a bona fide utterance $\boldsymbol{o}_{1:T, 1}$ of length $T$ and $K$ augmented utterances $\{\boldsymbol{o}_{1:T, 2},\cdots, \boldsymbol{o}_{1:T, K+1}\}$. 
Let us use $S$ vocoders to create spoofed vocoded data for the $1+K$ bona fide utterances, after which we get $S(1+K)$ vocoded utterances. 
With a CM feature extractor,  let ${\boldsymbol{x}}_{1:N, i}$ and $\tilde{\boldsymbol{x}}_{1:N, j}$ be the feature sequence extracted from $\boldsymbol{o}_{1:T, i}$ and $\tilde{\boldsymbol{o}}_{1:T, j}$, where $i \in \mathcal{I}=\{1, 2, \cdots, (1+K) \}$ and $j \in \mathcal{J}=\{1, 2, \cdots, S(1+K) \}$, respectively. 
If we create a mini-batch using ${\boldsymbol{x}}_{1:N, i}, \forall{i}\in\mathcal{I}$ and $\tilde{\boldsymbol{x}}_{1:N, j}$, $\forall{j}\in\mathcal{J}$, the contrastive feature loss over the mini-batch can be defined as
\begin{equation}
\begin{split}
%\mathcal{L}_{\text{CF}}  = &- \sum_{i\in\mathcal{I}} \frac{1}{|\mathcal{I}|-1} \sum_{p\in\mathcal{I}, p \neq i} \log \frac{\exp(f(\boldsymbol{x}_i, \boldsymbol{x}_p))}{\sum_{j\in\mathcal{J}}\exp(f(\boldsymbol{x}_i, \tilde{\boldsymbol{x}}_j))}  \\
%&- \sum_{j\in\mathcal{J}} \frac{1}{|\mathcal{J}|-1}  \sum_{q\in\mathcal{J}, q \neq j} \log \frac{\exp(f(\tilde{\boldsymbol{x}}_j, \tilde{\boldsymbol{x}}_q))}{\sum_{i\in\mathcal{I}}\exp(f(\tilde{\boldsymbol{x}}_j, {\boldsymbol{x}}_i))},
\mathcal{L}_{\text{CF}}  = &- \sum_{i\in\mathcal{I}} \frac{1}{|\mathcal{I}|-1} \sum_{p\in\mathcal{I}, p \neq i} \log \frac{\exp(f(\boldsymbol{x}_i, \boldsymbol{x}_p))}{\mathcal{H}(\boldsymbol{x}_i)}  \\
&- \sum_{j\in\mathcal{J}} \frac{1}{|\mathcal{J}|-1}  \sum_{q\in\mathcal{J}, q \neq j} \log \frac{\exp(f(\tilde{\boldsymbol{x}}_j, \tilde{\boldsymbol{x}}_q))}{\mathcal{H}(\tilde{\boldsymbol{x}}_j)},
\label{eq:loss}
\end{split}
\end{equation}
where $\mathcal{H}(\boldsymbol{z}) = \sum_{i=1}^{|\mathcal{I}|}\exp(f(\boldsymbol{z}, \boldsymbol{x}_i)) + \sum_{j=1}^{|\mathcal{J}|}\exp(f(\boldsymbol{z}, \tilde{\boldsymbol{x}}_j)) - \exp(f(\boldsymbol{z}, {\boldsymbol{z}}))$, $|\mathcal{I}| = K+1$, $|\mathcal{J}| = S(1+K)$, and where we drop the time sequence subscript ${}_{1:N}$ in $\boldsymbol{x}_i$ and $\tilde{\boldsymbol{x}}_j$.  The $f(\boldsymbol{x}_{1:N, i}, \tilde{\boldsymbol{x}}_{1:N, p}) = \frac{1}{N}\sum_{n}\frac{\boldsymbol{x}_{n, i}^\top\tilde{\boldsymbol{x}}_{n, p}}{\tau|| \boldsymbol{x}_{n, i} ||_2|| \tilde{\boldsymbol{x}}_{n, p} ||_2}$ measures the cosine similarity averaged over the feature vector pairs $\{\boldsymbol{x}_{n, i}, \tilde{\boldsymbol{x}}_{n, p}\}, \forall {n} \in\{1, \cdots, N\}$.  The hyper-parameter $\tau$ is fixed to 0.07 following \cite{khosla2020supervised}.

$\mathcal{L}_{\text{CF}}$ in Eq.~(\ref{eq:loss}) is a special case of supervised contrastive loss \cite{khosla2020supervised} that considers only two classes (i.e., \emph{bona fide} and \emph{spoofed}). A difference is that the similarity $f(\cdot)$ is computed between two aligned feature sequences, while it can be used for utterance-level features with $N=1$.
%Whenever $\mathcal{L}_{\text{CF}}$ is used in this study, it is directly merged with the conventional cross-entropy loss for CM training, i.e., $\mathcal{L}_{\text{CE}} + \mathcal{L}_{\text{CF}}$.
In this study, $\mathcal{L}_{\text{CF}}$ is directly merged with the conventional cross-entropy loss for CM training, i.e., $\mathcal{L}_{\text{CE}} + \mathcal{L}_{\text{CF}}$.
It is expected to drive the features of the data in the same class to be closer to each other while pushing away those from a different class.

\begin{table*}[!t]
\caption{CM training sets with vocoded spoofed data.  For reference, ASVspoof 2019 LA training set (\setTLAshort) contains 2,580 bona fide and 22,800 spoofed trials from 20 speakers at sampling rate of 16 kHz. Right-most column is sampling rate (SR) that vocoder works on.}
\label{tab:training_data}
\vspace{-7mm}
\begin{center}
\setlength{\tabcolsep}{3pt}
\resizebox{\textwidth}{!}{
\begin{tabular}{rcccllll}
\toprule
ID & \#. Spr. & \#. Bona. & \#. Spoof. & Vocoder type & Implementation & \shortstack{Vocoder train/fine-tune data}   & Vocoder SR\\
\midrule
\setTWFshort& \textcolor{white}{0}1 & 3,930 & 15,720  & HiFiGAN, MB-MelGAN, PWG, WaveGlow  & ESPNet toolkit   &  LJSpeech / -  & 24  kHz\\
\midrule
\setTVIshort & \multirow{4}{*}{\shortstack{20 \\ same as \\ \setTLAshort}}  & \multirow{4}{*}{2,580} & \multirow{4}{*}{10,320}   & HiFiGAN, MB-MelGAN, PWG, StyleMelGAN   & ESPNet toolkit  & LibriTTS /  - & 24 kHz\\
\setTVIIIshort   &  & & & HiFiGAN, NSF-HiFiGAN, Hn-NSF,  WaveGlow & in-house  & LibriTTS / -  & 16 kHz\\
\setTVIVshort  &   & &  & HiFiGAN, NSF-HiFiGAN, Hn-NSF,  WaveGlow & in-house & {\setTLAshort} bona. / -  & 16 kHz\\
\setTVVshort  & & &  &   HiFiGAN, NSF-HiFiGAN, Hn-NSF,  WaveGlow & in-house & LibriTTS /  {\setTLAshort} bona. & 16  kHz\\
\bottomrule
\end{tabular}
}
\vspace{-5mm}
\end{center}
\label{tab:tablesets}
\end{table*}

\section{Experiments}
\label{sec:exp}

The first part of the experiments compares the training sets created in Section~\ref{sec:database} with a standard training set. The goal is to identify factors that one should be alerted to when creating vocoded spoofed data. The second part examines the performance of the contrastive feature loss introduced in Section~\ref{sec:sup}. With the experimental results, we give tentative answers to the questions raised in Introduction.

\subsection{Datasets}
\label{sec:database}

The CM training sets used in the experiments included those listed in Table~\ref{tab:training_data} and the commonly used ASVspoof 2019 LA training set (\setTLAshort) for reference. The development set was the ASVspoof 2019 LA development set no matter which training set was used. 

Eight test sets were used to fairly measure the CM performance. They included the test sets from ASVspoof 2019 LA  (\setELA), the evaluation subsets of ASVspoof 2021 LA (\setELAII), and the 2021 DF (\setEDF) tasks. 
Since CMs may overfit to the length of non-speech segments \cite{LiuASVspoof2021, muller21_asvspoof}, we included another version of the three test sets in which the non-speech segments were trimmed. Two of them ({\setELAHID} and {\setEDFHID}) are the official hidden subsets in the ASVspoof 2021 LA and DF datasets \cite{LiuASVspoof2021}, and the other one, {\setELATRIM}, was created by us given {\setELA}\footnote{Information on the ASVspoof2021 official hidden subsets is hosted on https://github.com/asvspoof-challenge/2021. The unofficial {\setELATRIM} and code are hosted on https://github.com/nii-yamagishilab/project-NN-Pytorch-scripts/tree/master/project/09-asvspoof-vocoded-trn.}.

The last two test sets were the entire WaveFake (\setEWFE) and In the Wild datasets (\setEDFE)\cite{muller2022does}. Note that, since the training set {\setTWFshort} is a subset of WaveFake, we excluded the result if a CM was trained on {\setTWFshort} and tested on {\setEWFE}. We used the entire WaveFake as a test set so that the results can be compared with other literature.  Note that WaveFake contains three more vocoders not covered by any training data in Table~\ref{tab:training_data}. 
The {\setEDFE} is a recently released dataset containing bona fide and spoofed trials from more than 50 English-speaking celebrities and politicians. The data has more diverse acoustic characteristics and is more challenging \cite{muller2022does}.

\subsection{Training Recipe and Evaluation Metric}

The same CM architecture was used in all the experiments: a two-staged model with a self-supervised-learning (SSL)-based feature extractor and a feedforward classifier. The SSL feature extractor was a pre-trained Wav2vec 2.0 model \cite{NEURIPS2020_92d1e1eb}\footnote{Model XLSR-53 at https://github.com/facebookresearch/fairseq/ blob/main/examples/wav2vec/README.md}. 
The classifier consisted of a global average pooling layer, a neural network with three fully-connected layers and LeakyReLU activation functions, and a linear output layer for binary classification. The weights of the pre-trained SSL feature extractor were updated during CM training. 
Similar SSL-based CMs have shown better generalization performance \cite{wang2021investigating, tak2022automatic}. 
Additional experiments were also conducted on two other CMs without a SSL feature extractor, but the results were not satisfying. Those additional results are uploaded to Arxiv\footnote{Arxiv Link: http://arxiv.org/abs/2210.10570}.

The CMs used the Adam optimizer ($\beta_1=0.9, \beta_2=0.999, \epsilon=10^{-8}$) \cite{kingma2014adam}. The learning rate was initialized to $5\times10^{-6}$ and multiplied by $0.1$ every ten epochs. The training process was stopped when the loss on the development set did not improve within ten epochs. The batch size was eight, but when the contrastive feature loss was used, each mini-batch contained one bona fide utterance and its four vocoded versions. 
The training data were truncated into segments with a maximum duration of 4s to fit the GPU memory. During inference, the whole trial was processed without truncation.

CM performance was measured using the equal error rate (EER). 
The EER was calculated on each individual test set and the pooled test set. 
Each CM was trained and evaluated for three rounds, where each round used a different random seed to initialize the network weights. The EER averaged over the three rounds is reported\footnote{ 
Statistical analysis on EERs \cite{wang2021comparative} was conducted. See results on Arxiv. }.

\begin{table}[!t]
\caption{Test set EERs (\%) of Experiment 1. Each column corresponds to results of CM trained on one training set. Darker cell color indicates higher EER value. Some results in column {\setTWFshort} are removed due to data overlapping (see Section~\ref{sec:database}).}
\vspace{-6mm}
\begin{center}
\footnotesize
\setlength{\tabcolsep}{4pt}
\resizebox{\columnwidth}{!}
{
\begin{tabular}{clcccccc}
\toprule
  &   &  \setTLA  &  \setTWF  &  \setTVI  & \setTVIII & \setTVIV  &  \setTVV \\ 
\midrule
\multirow{11}{*}{\rotatebox{90}{Test sets}}     &   \setELA   & \cellcolor[rgb]{0.99, 0.99, 0.99} 2.98 & \cellcolor[rgb]{0.66, 0.66, 0.66} 44.48 & \cellcolor[rgb]{0.97, 0.97, 0.97} 5.78 & \cellcolor[rgb]{0.98, 0.98, 0.98} 5.32 & \cellcolor[rgb]{0.96, 0.96, 0.96} 8.74 & \cellcolor[rgb]{0.98, 0.98, 0.98} 4.36\\ 
  &  \setELAII  & \cellcolor[rgb]{0.96, 0.96, 0.96} 7.53 & \cellcolor[rgb]{0.69, 0.69, 0.69} 41.57 & \cellcolor[rgb]{0.84, 0.84, 0.84} 26.30 & \cellcolor[rgb]{0.90, 0.90, 0.90} 17.98 & \cellcolor[rgb]{0.89, 0.89, 0.89} 19.29 & \cellcolor[rgb]{0.86, 0.86, 0.86} 24.39\\ 
  &   \setEDF   & \cellcolor[rgb]{0.97, 0.97, 0.97} 6.67 & \cellcolor[rgb]{0.86, 0.86, 0.86} 24.26 & \cellcolor[rgb]{0.94, 0.94, 0.94} 11.95 & \cellcolor[rgb]{0.95, 0.95, 0.95} 11.54 & \cellcolor[rgb]{0.96, 0.96, 0.96} 9.71 & \cellcolor[rgb]{0.94, 0.94, 0.94} 13.31\\ 
\cmidrule{2-8}
  & \setELATRIM & \cellcolor[rgb]{0.92, 0.92, 0.92} 15.56 & \cellcolor[rgb]{0.80, 0.80, 0.80} 31.62 & \cellcolor[rgb]{0.86, 0.86, 0.86} 23.29 & \cellcolor[rgb]{0.92, 0.92, 0.92} 16.16 & \cellcolor[rgb]{0.92, 0.92, 0.92} 14.99 & \cellcolor[rgb]{0.96, 0.96, 0.96} 9.52\\ 
  & \setELAHID  & \cellcolor[rgb]{0.82, 0.82, 0.82} 28.80 & \cellcolor[rgb]{0.83, 0.83, 0.83} 27.60 & \cellcolor[rgb]{0.82, 0.82, 0.82} 28.30 & \cellcolor[rgb]{0.89, 0.89, 0.89} 19.49 & \cellcolor[rgb]{0.90, 0.90, 0.90} 17.62 & \cellcolor[rgb]{0.88, 0.88, 0.88} 21.43\\ 
  & \setEDFHID  & \cellcolor[rgb]{0.86, 0.86, 0.86} 23.62 & \cellcolor[rgb]{0.84, 0.84, 0.84} 26.18 & \cellcolor[rgb]{0.87, 0.87, 0.87} 22.01 & \cellcolor[rgb]{0.93, 0.93, 0.93} 13.92 & \cellcolor[rgb]{0.94, 0.94, 0.94} 13.50 & \cellcolor[rgb]{0.91, 0.91, 0.91} 16.99\\ 
  &  \setEWFE   & \cellcolor[rgb]{0.92, 0.92, 0.92} 15.76 & - & \cellcolor[rgb]{0.72, 0.72, 0.72} 39.27 & \cellcolor[rgb]{0.77, 0.77, 0.77} 34.05 & \cellcolor[rgb]{0.91, 0.91, 0.91} 17.10 & \cellcolor[rgb]{0.95, 0.95, 0.95} 10.89\\ 
  &  \setEDFE   & \cellcolor[rgb]{0.84, 0.84, 0.84} 26.65 & \cellcolor[rgb]{0.89, 0.89, 0.89} 19.98 & \cellcolor[rgb]{0.70, 0.70, 0.70} 41.06 & \cellcolor[rgb]{0.75, 0.75, 0.75} 36.46 & \cellcolor[rgb]{0.87, 0.87, 0.87} 22.26 & \cellcolor[rgb]{0.89, 0.89, 0.89} 19.45\\ 
\cmidrule{2-8}
  &  \setGAll   & \cellcolor[rgb]{0.93, 0.93, 0.93} 14.24 & - & \cellcolor[rgb]{0.75, 0.75, 0.75} 36.57 & \cellcolor[rgb]{0.71, 0.71, 0.71} 39.95 & \cellcolor[rgb]{0.89, 0.89, 0.89} 19.39 & \cellcolor[rgb]{0.92, 0.92, 0.92} 16.35\\ 
\bottomrule
\end{tabular}
}
\vspace{-7mm}
\end{center}
\label{tab:exp-1}
\end{table}

\subsection{Experiment 1: Comparing Vocoded Training Sets}
\label{sec:exp1}

This experiment investigated the effectiveness of the training sets with vocoded spoofed data listed in Table~\ref{tab:tablesets}. The  {\setTLAshort} and  {\setTWFshort} were also included for reference. The CM was trained on each training set with the standard binary CE loss $\mathcal{L}_{\text{CE}}$.  From the results listed in Table~\ref{tab:exp-1},  we first note that the EERs given {\setTLAshort} were similar to those reported in \cite{wang2021investigating}. This indicates that the CM and training recipe worked as expected.  

Compared with the standard training set {\setTLAshort}, the EERs when using vocoded training sets were higher on the first three test sets. 
%However, this may be because the CM trained on {\setTLAshort} exploited the spurious information on the length of non-speech segments \cite{LiuASVspoof2021, muller21_asvspoof}. 
However, this may be because the CM trained on {\setTLAshort} exploited non-speech segments that unevenly distributed in bona fide and spoofed trials \cite{LiuASVspoof2021, muller21_asvspoof}.
On other test sets, the CM trained on {\setTVVshort} achieved lower EERs.
%The other training sets listed in Table~\ref{tab:tablesets} led to varied performance, but 
The results obtained on the vocoded training sets allowed us to answer \textbf{whether there is any caveat when creating useful vocoded spoofed data}. Described below are our tentative suggestions: 

\emph{Avoid waveform re-sampling during copy-synthesis}: 
%A comparison between {\setTVIshort} and {\setTVIIshort} suggests that it is better to use neural vocoders with a compatible sampling rate. k
Although the vocoders in {\setTVIshort} and {\setTVIIshort} are not the same, the larger EERs in {\setTVIshort} may be partially caused by re-sampling.  This hypothesis was supported by another experiment using the HiFiGAN models operating at 24 and 16 kHz (see Appendix).
Neural vocoders, which have never been trained on re-sampled waveforms, may produce artifacts in vocoded data if the input bona fide data is re-sampled. CMs that overfit to those artifacts may not generalize to realistic spoofing attacks that do not do re-sampling.

\emph{Tune neural vocoders to target domain}: From comparisons among {\setTVIIshort}, {\setTVIVshort}, and {\setTVVshort}, we observe that it is preferable to let neural vocoders learn from the bona fide data from which the vocoded spoofed data are to be created.  This can be done by training the neural vocoders from scratch or fine-tuning pre-trained neural vocoders using the bona fide data. 
The {\setTWFshort} training set did not perform well probably because of the domain mismatch and limited number of speakers.

\begin{table}[!t]
\caption{Test set EERs (\%) of Experiment 2. Each column corresponds to CM trained using specific configuration. Darker cell color indicates higher EER value.  Results in columns \selfcircle{1} and \selfcircle{2} are copied from Table~\ref{tab:exp-1} as reference.}
\vspace{-6mm}
\begin{center}
\setlength{\tabcolsep}{2pt}
\resizebox{\columnwidth}{!}
{
\begin{tabular}{clccccccc}
\toprule
  &   \multicolumn{1}{l}{Training criterion}  & \multicolumn{4}{c}{$\mathcal{L}_{\text{CE}}$ }   & \multicolumn{3}{c}{$\mathcal{L}_{\text{CE}} + \mathcal{L}_{\text{CF}}$ }   \\
  \cmidrule(lr){3-6}\cmidrule(lr){7-9} 
   &  \multicolumn{1}{l}{Data augmentation} &  \multicolumn{2}{c}{$\times$} & \multicolumn{2}{c}{RawBoost} &  \multicolumn{3}{c}{RawBoost}  \\
  \cmidrule(lr){3-4}\cmidrule(lr){5-6}\cmidrule(lr){7-9} 
  &   \multicolumn{1}{l}{Training set}     & \setTLA    & \setTVV & \setTLA & \setTVV & \setTLA & \setTVV & \setTVV\\ 
  & \multicolumn{1}{l}{Bona-spoof paired}   &  $\times$  &  $\times$  &  $\times$  &  $\times$  &  $\times$  &  $\times$ &  $\checkmark$ \\
 & \multicolumn{1}{l}{ID}  & \selfcircle{1} & \selfcircle{2} & \selfcircle{3} & \selfcircle{4} & \selfcircle{5} & \selfcircle{6} & \selfcircle{7} \\
\midrule
\multirow{11}{*}{\rotatebox{90}{Test sets}} &   \setELA   & \cellcolor[rgb]{0.99, 0.99, 0.99} 2.98 & \cellcolor[rgb]{0.99, 0.99, 0.99} 4.36 & \cellcolor[rgb]{1.00, 1.00, 1.00} 0.22 & \cellcolor[rgb]{0.99, 0.99, 0.99} 3.46 & \cellcolor[rgb]{1.00, 1.00, 1.00} 0.21 & \cellcolor[rgb]{1.00, 1.00, 1.00} 2.63 & \cellcolor[rgb]{1.00, 1.00, 1.00} 2.21\\ 
  &  \setELAII  & \cellcolor[rgb]{0.97, 0.97, 0.97} 7.53 & \cellcolor[rgb]{0.81, 0.81, 0.81} 24.39 & \cellcolor[rgb]{0.99, 0.99, 0.99} 3.63 & \cellcolor[rgb]{0.89, 0.89, 0.89} 16.55 & \cellcolor[rgb]{0.99, 0.99, 0.99} 3.30 & \cellcolor[rgb]{0.89, 0.89, 0.89} 16.67 & \cellcolor[rgb]{0.88, 0.88, 0.88} 17.90\\ 
  &   \setEDF   & \cellcolor[rgb]{0.97, 0.97, 0.97} 6.67 & \cellcolor[rgb]{0.92, 0.92, 0.92} 13.31 & \cellcolor[rgb]{0.99, 0.99, 0.99} 3.65 & \cellcolor[rgb]{0.95, 0.95, 0.95} 9.60 & \cellcolor[rgb]{0.99, 0.99, 0.99} 4.12 & \cellcolor[rgb]{0.97, 0.97, 0.97} 6.92 & \cellcolor[rgb]{0.98, 0.98, 0.98} 5.04\\ 
  \cmidrule{2-9}
  & \setELATRIM & \cellcolor[rgb]{0.90, 0.90, 0.90} 15.56 & \cellcolor[rgb]{0.95, 0.95, 0.95} 9.52 & \cellcolor[rgb]{0.96, 0.96, 0.96} 9.16 & \cellcolor[rgb]{0.98, 0.98, 0.98} 6.09 & \cellcolor[rgb]{0.96, 0.96, 0.96} 9.00 & \cellcolor[rgb]{0.99, 0.99, 0.99} 4.48 & \cellcolor[rgb]{0.99, 0.99, 0.99} 3.79\\ 
  & \setELAHID  & \cellcolor[rgb]{0.76, 0.76, 0.76} 28.80 & \cellcolor[rgb]{0.85, 0.85, 0.85} 21.43 & \cellcolor[rgb]{0.85, 0.85, 0.85} 21.18 & \cellcolor[rgb]{0.87, 0.87, 0.87} 19.37 & \cellcolor[rgb]{0.78, 0.78, 0.78} 26.98 & \cellcolor[rgb]{0.91, 0.91, 0.91} 15.05 & \cellcolor[rgb]{0.91, 0.91, 0.91} 14.57\\ 
  & \setEDFHID  & \cellcolor[rgb]{0.82, 0.82, 0.82} 23.62 & \cellcolor[rgb]{0.89, 0.89, 0.89} 16.99 & \cellcolor[rgb]{0.92, 0.92, 0.92} 13.64 & \cellcolor[rgb]{0.92, 0.92, 0.92} 14.29 & \cellcolor[rgb]{0.89, 0.89, 0.89} 16.85 & \cellcolor[rgb]{0.96, 0.96, 0.96} 8.17 & \cellcolor[rgb]{0.96, 0.96, 0.96} 7.78\\ 
  &  \setEWFE   & \cellcolor[rgb]{0.90, 0.90, 0.90} 15.76 & \cellcolor[rgb]{0.95, 0.95, 0.95} 10.89 & \cellcolor[rgb]{0.79, 0.79, 0.79} 26.37 & \cellcolor[rgb]{0.97, 0.97, 0.97} 6.87 & \cellcolor[rgb]{0.81, 0.81, 0.81} 24.62 & \cellcolor[rgb]{0.99, 0.99, 0.99} 4.03 & \cellcolor[rgb]{1.00, 1.00, 1.00} 2.50\\ 
  &  \setEDFE   & \cellcolor[rgb]{0.78, 0.78, 0.78} 26.65 & \cellcolor[rgb]{0.87, 0.87, 0.87} 19.45 & \cellcolor[rgb]{0.90, 0.90, 0.90} 16.17 & \cellcolor[rgb]{0.94, 0.94, 0.94} 12.08 & \cellcolor[rgb]{0.89, 0.89, 0.89} 17.07 & \cellcolor[rgb]{0.96, 0.96, 0.96} 9.37 & \cellcolor[rgb]{0.97, 0.97, 0.97} 7.55\\ 
    \cmidrule{2-9}
  &  \setGAll   & \cellcolor[rgb]{0.92, 0.92, 0.92} 14.24 & \cellcolor[rgb]{0.90, 0.90, 0.90} 16.35 & \cellcolor[rgb]{0.93, 0.93, 0.93} 13.12 & \cellcolor[rgb]{0.93, 0.93, 0.93} 13.13 & \cellcolor[rgb]{0.92, 0.92, 0.92} 13.68 & \cellcolor[rgb]{0.93, 0.93, 0.93} 13.15 & \cellcolor[rgb]{0.94, 0.94, 0.94} 11.27\\ 
\bottomrule
\end{tabular}
}
\vspace{-5mm}
\end{center}
\label{tab:exp-2}
\end{table}

\subsection{Experiment 2: Effect of Contrastive Feature Loss}
\label{sec:exp2}

With the promising result of {\setTVVshort}  in Experiment 1, we investigated \textbf{whether the CM performance can be further improved if the vocoded spoofed data can be put to good use}. The method examined was the contrastive feature loss ($\mathcal{L}_{\text{CF}}$ in Eq.~(\ref{eq:loss})), which was computed on both the SSL-based front-end's output sequence and the utterance-level vector after global average pooling.
We trained a few CMs with different configurations listed in Table~\ref{tab:exp-2}. 
Marked by  $\selfcircle{7}$ is the experimental CM trained using $\mathcal{L}_{\text{CF}}+\mathcal{L}_{\text{CE}}$. 
Each mini-batch for $\selfcircle{7}$ contained a bona fide trial, its vocoded spoofed data from the four neural vocoders, and one piece of augmented data for each trial (i.e., $S=4$ and $K=1$ for $\mathcal{L}_{\text{CF}}$). The data augmentation was done using RawBoost \cite{Tak2021}.

For comparison, CM $\selfcircle{6}$ used the same setting as $\selfcircle{7}$ except that the spoofed trial in each mini-batch was randomly selected and was not paired with the bona fide trial. $\selfcircle{5}$ was the same as $\selfcircle{6}$, but the training data was from {\setTLAshort}. Note that the mini-batch cannot include paired bona fide and vocoded spoofed trials when using {\setTLAshort}.   $\selfcircle{4}$ and $\selfcircle{3}$ further removed $\mathcal{L}_{\text{CF}}$.  Finally, $\selfcircle{2}$ and $\selfcircle{1}$ are from Experiment 1 and were included as reference.

Results suggest that \emph{the contrastive feature loss can improve the CM when it is used together with a training set that contains vocoded spoofed data}. 
From \selfcircle{2} to \selfcircle{4} and then \selfcircle{7}, we observe that the overall pooled EER decreased from 16.35\% to 11.27\%.
While the comparison between \selfcircle{2} and \selfcircle{4} suggests that data augmentation such as RawBoost is helpful, the comparison between \selfcircle{4}  and \selfcircle{7} implies that adding the contrastive feature loss brings in additional improvement. Furthermore, a comparison between \selfcircle{6}  and \selfcircle{7} suggests that using a bona fide trial and its corresponding vocoded spoofed data in each mini-batch is also helpful. 

The gain brought by the contrastive feature loss diminishes when using {\setTLAshort}. This can be observed from the comparison between \selfcircle{3} and \selfcircle{5}. Hence, it is recommended to add the contrastive feature loss when using a training set with vocoded spoofed data.

\subsection{Analysis and Comparison with Other Literature}
\label{sec:exp3}
\begin{figure}[t!]
\centering
\includegraphics[trim=0 20 0 10, width=\columnwidth]{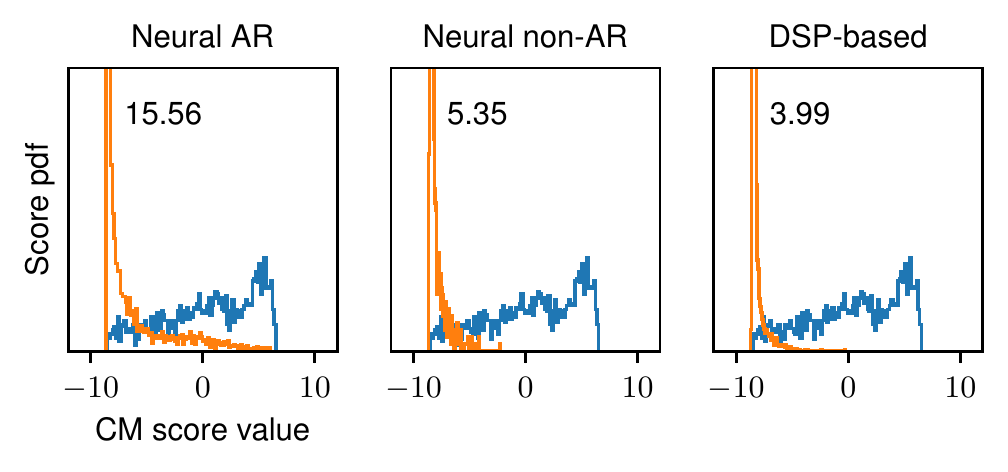}
\vspace{-2mm}
\caption{Score distributions of \selfcircle{7} on \textcolor{blue}{bona fide} and \textcolor{orange}{spoofed} trials in {\setEDFHID}. Number in each sub-figure is EER (\%).} 
\label{fig:scoredist}
\end{figure}

Since CMs trained on {\setTVVshort} were exposed only to a limited number of neural vocoders, \textbf{do they generalize to TTS and VC spoofing attacks using unseen vocoders?} 
To answer this question, we followed \cite{LiuASVspoof2021} and analyzed the score distributions after grouping the trials in {\setEDFHID} according to the category of the vocoder used in the spoofing TTS and VC systems. The three categories were neural AR, neural non-AR, and traditional DSP-based vocoders. Scores given by CM \selfcircle{7} are plotted in Fig.~\ref{fig:scoredist}.  As expected, the CM performed worse on spoofed trials with neural AR vocoders since the CM training data does not cover similar vocoders. While DSP-based vocoders are also unseen, the CM trained on {\setTVVshort} seemed to be generalizable. Similar results were observed on other test sets that had labels of spoofing attacks (see results on Arxiv).
In short, \emph{a CM trained on bona fide and vocoded spoofed data may not perfectly generalize to spoofing attacks from unseen vocoder types}.

Despite the limitation, we highlight that \emph{the best CM \selfcircle{7} in this study has demonstrated competitive performance in detecting actual TTS/VC attacks}. From the comparison with results in other literature listed in Table~\ref{tab:eer_comp}, it is clear that the EER of \selfcircle{7} outperformed the others on all the test sets except {\setELA} and {\setELAII}.  
The unsatisfactory performance on {\setELAII}  may be alleviated if we use a codec for data augmentation. This is left for future work.

\begin{table}[t!]
\caption{Comparison of EERs (\%) with other literature. Bolded \textbf{EER} marks statistically significantly best value. On LA and DF test sets, we include only CMs that have results available on \emph{both} original and non-speech-trimmed data. \cite{muller21_asvspoof} used different tool to trim non-speech in {\setELATRIM}, but results can be compared.
}
\vspace{-6mm}
\begin{center}
\footnotesize
\setlength{\tabcolsep}{2pt}{
\begin{tabular}{lrr}
\toprule
 &  {\setELA} & {\setELATRIM} \\ 
 \midrule
\cite{muller21_asvspoof} & 7.35  & *35.32 \\
Appendix & \textbf{1.54} & 17.74 \\
Ours &  2.21 & \textbf{3.79} \\ 
 \bottomrule
\end{tabular}
\quad
\begin{tabular}{lrr}
\toprule
 & {\setELAII} & {\setELAHID} \\ 
 \midrule
 \cite{LiuASVspoof2021}& \textbf{1.32} & $>$22  \\
\cite{LiuASVspoof2021} & 2.77 & $>$22 \\
Ours & 17.9  &  \textbf{14.57} \\ 
 \bottomrule
\end{tabular}\quad
\begin{tabular}{lrr}
\\
\toprule
 & {\setEDF} & {\setEDFHID} \\ 
 \midrule
 \cite{LiuASVspoof2021}& 15.64 & $>$20  \\
\cite{LiuASVspoof2021} & 16.05 & $>$20 \\
Ours & \textbf{5.04}  &  \textbf{7.78} \\ 
 \bottomrule
\end{tabular}\quad
\begin{tabular}{lr}
\\
\toprule
 & {\setEWFE}\\ 
 \midrule
 \cite{Wang2023}& 8.29 \\
Ours & \textbf{2.50}  \\ 
\\
 \bottomrule
\end{tabular}\quad
\begin{tabular}{lr}
\\
\toprule
 & {\setEDFE}\\ 
 \midrule
 \cite{pianese2022deepfake}& 15.00 \\
  \cite{muller2022does}& 33.00 \\
Ours & \textbf{7.55}  \\ 
 \bottomrule
\end{tabular}
}
\vspace{-5mm}
\end{center}
\label{tab:eer_comp}
\end{table}

\section{Conclusion}
\label{sec:con}
This study investigated an alternative way to create a training set for spoofing CM. Instead of building full-fledged TTS and VC systems, this study simply created spoofed data by conducting copy-synthesis on bona fide utterances using neural vocoders. Multiple training sets were created on the basis of the ASVspoof 2019 LA training set and compared in the experiments. Furthermore, this study introduced a contrastive feature loss to better use the bona fide trials and their corresponding vocoded spoofed data. 

The experiments produced a few useful findings. First, it is preferable to use neural vocoders that are matched with the bona fide data in terms of the sampling rate and acoustic conditions; the latter can be achieved by training or fine-tuning neural vocoders using the bona fide data to be vocoded. Second, when such a training set is used, the contrastive feature loss with a special mini-match pairing strategy can improve the CM performance. Third, the trained CM showed promising generalizability to multiple challenging test sets, even though its performance degraded on spoofing attacks generated by unseen neural vocoders as expected. We hope that these findings can encourage practitioners to build their own training sets.

\vfill\pagebreak


\pretolerance=1000

\bibliographystyle{IEEEbib}
\bibliography{library}

\begin{thebibliography}{1}

\bibitem{wang2021comparative}
Xin Wang and Junichi Yamagishi,
\newblock ``{A comparative study on recent neural spoofing countermeasures for
  synthetic speech detection},''
\newblock in {\em Proc. Interspeech}, 2021, pp. 4259--4263.

\bibitem{jung2022aasist}
Jee-weon Jung, Hee-Soo Heo, Hemlata Tak, Hye-jin Shim, Joon~Son Chung, Bong-Jin
  Lee, Ha-Jin Yu, and Nicholas Evans,
\newblock ``{AASIST: Audio anti-spoofing using integrated spectro-temporal
  graph attention networks},''
\newblock in {\em Proc. ICASSP}. IEEE, 2022, pp. 6367--6371.

\bibitem{Wang2022}
Xin Wang and Junichi Yamagishi,
\newblock ``{A Practical Guide to Logical Access Voice Presentation Attack
  Detection},''
\newblock in {\em Frontiers in Fake Media Generation and Detection}, pp.
  169--214. Springer, 2022.

\bibitem{wang2021investigating}
Xin Wang and Junichi Yamagishi,
\newblock ``{Investigating Self-Supervised Front Ends for Speech Spoofing
  Countermeasures},''
\newblock in {\em Proc. Odyssey}, 2022, pp. 100--106.

\bibitem{tak2022automatic}
Hemlata Tak, Massimiliano Todisco, Xin Wang, Jee-weon Jung, Junichi Yamagishi,
  and Nicholas Evans,
\newblock ``{Automatic speaker verification spoofing and deepfake detection
  using wav2vec 2.0 and data augmentation},''
\newblock in {\em Proc. Odyssey}, 2022, pp. 112--119.

\bibitem{wu2015asvspoof}
Zhizheng Wu, Tomi Kinnunen, Nicholas Evans, Junichi Yamagishi, Cemal
  Hanil{\c{c}}i, Md~Sahidullah, and Aleksandr Sizov,
\newblock ``{ASVspoof 2015: the first automatic speaker verification spoofing
  and countermeasures challenge},''
\newblock in {\em Proc. Interspeech}, 2015, pp. 2037--2041.

\end{thebibliography}


\begin{thebibliography}{10}

\bibitem{evans2013spoofing}
Nicholas Evans, Tomi Kinnunen, and Junichi Yamagishi,
\newblock ``{Spoofing and countermeasures for automatic speaker
  verification},''
\newblock in {\em Proc. Interspeech}, 2013, pp. 925--929.

\bibitem{LiuASVspoof2021}
Xuechen Liu, Xin Wang, Md~Sahidullah, Jose Patino, H{\'{e}}ctor Delgado, Tomi
  Kinnunen, Massimiliano Todisco, Junichi Yamagishi, Nicholas Evans, Andreas
  Nautsch, and Kong~Aik Lee,
\newblock ``{ASVspoof 2021: Towards Spoofed and Deepfake Speech Detection in
  the Wild},''
\newblock {\em arXiv preprint arXiv:2210.02437}, 2022.

\bibitem{asvspoof2019_database}
Xin Wang, Junichi Yamagishi, Massimiliano Todisco, and Others,
\newblock ``{ASVspoof 2019: A large-scale public database of synthesized,
  converted and replayed speech},''
\newblock {\em Computer Speech \& Language}, vol. 64, pp. 101114, nov 2020.

\bibitem{cooper2020zero}
Erica Cooper, Cheng-I Lai, Yusuke Yasuda, Fuming Fang, Xin Wang, Nanxin Chen,
  and Junichi Yamagishi,
\newblock ``{Zero-shot multi-speaker text-to-speech with state-of-the-art
  neural speaker embeddings},''
\newblock in {\em Proc. ICASSP}, 2020, pp. 6184--6188.

\bibitem{casanova2022yourtts}
Edresson Casanova, Julian Weber, Christopher~D Shulby, Arnaldo~Candido Junior,
  Eren G{\"{o}}lge, and Moacir~A Ponti,
\newblock ``{Yourtts: Towards zero-shot multi-speaker TTS and zero-shot voice
  conversion for everyone},''
\newblock in {\em Proc. ICML}, 2022, pp. 2709--2720.

\bibitem{wu2013synthetic}
Zhizheng Wu, Xiong Xiao, Eng~Siong Chng, and Haizhou Li,
\newblock ``{Synthetic speech detection using temporal modulation feature},''
\newblock in {\em Proc. ICASSP}, 2013, pp. 7234--7238.

\bibitem{sanchez2014cross}
Jon Sanchez, Ibon Saratxaga, Inma Hernaez, Eva Navas, and Daniel Erro,
\newblock ``{A cross-vocoder study of speaker independent synthetic speech
  detection using phase information},''
\newblock in {\em Proc. Interspeech}, 2014.

\bibitem{sizov2015joint}
Aleksandr Sizov, Elie Khoury, Tomi Kinnunen, Zhizheng Wu, and S{\'{e}}bastien
  Marcel,
\newblock ``{Joint speaker verification and antispoofing in the i-vector
  space},''
\newblock {\em IEEE Transactions on Information Forensics and Security}, vol.
  10, no. 4, pp. 821--832, 2015.

\bibitem{frank2021wavefake}
Joel Frank and Lea Sch{\"{o}}nherr,
\newblock ``{WaveFake: A Data Set to Facilitate Audio DeepFake Detection},''
\newblock in {\em Proc. NeurIPS Datasets and Benchmarks 2021}, 2021.

\bibitem{oord2016wavenet}
Aaron van~den Oord, Sander Dieleman, Heiga Zen, Karen Simonyan, Oriol Vinyals,
  Alex Graves, Nal Kalchbrenner, Andrew Senior, and Koray Kavukcuoglu,
\newblock ``{WaveNet: A generative model for raw audio},''
\newblock in {\em Proc. SSW}, 2016.

\bibitem{yamamoto2020parallel}
Ryuichi Yamamoto, Eunwoo Song, and Jae-Min Kim,
\newblock ``{Parallel WaveGAN: A fast waveform generation model based on
  generative adversarial networks with multi-resolution spectrogram},''
\newblock in {\em Proc. ICASSP}. IEEE, 2020, pp. 6199--6203.

\bibitem{NEURIPS2020_c5d73680}
Jungil Kong, Jaehyeon Kim, and Jaekyoung Bae,
\newblock ``{HiFi-GAN: Generative Adversarial Networks for Efficient and High
  Fidelity Speech Synthesis},''
\newblock in {\em Proc. NIPS}, 2020, vol.~33, pp. 17022--17033.

\bibitem{yang2021multi}
Geng Yang, Shan Yang, Kai Liu, Peng Fang, Wei Chen, and Lei Xie,
\newblock ``{Multi-band MelGAN: Faster waveform generation for high-quality
  text-to-speech},''
\newblock in {\em Proc. SLT}, 2021, pp. 492--498.

\bibitem{wangNSFall}
Xin Wang, Shinji Takaki, and Junichi Yamagishi,
\newblock ``{Neural Source-Filter Waveform Models for Statistical Parametric
  Speech Synthesis},''
\newblock {\em IEEE/ACM Transactions on Audio, Speech, and Language
  Processing}, vol. 28, pp. 402--415, 2020.

\bibitem{tomashenko2022voiceprivacy}
Natalia Tomashenko, Xin Wang, Xiaoxiao Miao, Hubert Nourtel, Pierre Champion,
  Massimiliano Todisco, Emmanuel Vincent, Nicholas Evans, Junichi Yamagishi,
  and Jean~Fran{\c{c}}ois Bonastre,
\newblock ``{The VoicePrivacy 2022 Challenge Evaluation Plan},''
\newblock {\em arXiv preprint arXiv:2203.12468}, 2022.

\bibitem{prenger2018waveglow}
Ryan Prenger, Rafael Valle, and Bryan Catanzaro,
\newblock ``{WaveGlow: A Flow-based Generative Network for Speech Synthesis},''
\newblock in {\em Proc. ICASSP}, 2019, pp. 3617--3621.

\bibitem{hayashi2020espnet}
Tomoki Hayashi, Ryuichi Yamamoto, Katsuki Inoue, Takenori Yoshimura, Shinji
  Watanabe, Tomoki Toda, Kazuya Takeda, Yu~Zhang, and Xu~Tan,
\newblock ``{Espnet-TTS: Unified, reproducible, and integratable open source
  end-to-end text-to-speech toolkit},''
\newblock in {\em Proc. ICASSP}, 2020, pp. 7654--7658.

\bibitem{zen2019libritts}
Heiga Zen, Viet Dang, Rob Clark, Yu~Zhang, Ron~J Weiss, Ye~Jia, Zhifeng Chen,
  and Yonghui Wu,
\newblock ``{LibriTTS: A Corpus Derived from LibriSpeech for Text-to-Speech},''
\newblock in {\em Proc. Interspeech}, sep 2019, pp. 1526--1530.

\bibitem{muller21_asvspoof}
Nicolas M{\"{u}}ller, Franziska Dieckmann, Pavel Czempin, Roman Canals,
  Konstantin B{\"{o}}ttinger, and Jennifer Williams,
\newblock ``{Speech is Silver, Silence is Golden: What do ASVspoof-trained
  Models Really Learn?},''
\newblock in {\em Proc. ASVspoof Challenge workshop}, 2021, pp. 55--60.

\bibitem{khosla2020supervised}
Prannay Khosla, Piotr Teterwak, Chen Wang, Aaron Sarna, Yonglong Tian, Phillip
  Isola, Aaron Maschinot, Ce~Liu, and Dilip Krishnan,
\newblock ``{Supervised contrastive learning},''
\newblock in {\em Proc. NIPS}, 2020, pp. 18661--18673.

\bibitem{muller2022does}
Nicolas~M M{\"{u}}ller, Pavel Czempin, Franziska Dieckmann, Adam Froghyar, and
  Konstantin B{\"{o}}ttinger,
\newblock ``{Does Audio Deepfake Detection Generalize?},''
\newblock {\em Proc. Interspeech}, pp. 2783--2787, 2022.

\bibitem{NEURIPS2020_92d1e1eb}
Alexei Baevski, Yuhao Zhou, Abdelrahman Mohamed, and Michael Auli,
\newblock ``{wav2vec 2.0: A Framework for Self-Supervised Learning of Speech
  Representations},''
\newblock in {\em Proc. NIPS}, 2020, vol.~33, pp. 12449--12460.

\bibitem{wang2021investigating}
Xin Wang and Junichi Yamagishi,
\newblock ``{Investigating Self-Supervised Front Ends for Speech Spoofing
  Countermeasures},''
\newblock in {\em Proc. Odyssey}, 2022, pp. 100--106.

\bibitem{tak2022automatic}
Hemlata Tak, Massimiliano Todisco, Xin Wang, Jee-weon Jung, Junichi Yamagishi,
  and Nicholas Evans,
\newblock ``{Automatic speaker verification spoofing and deepfake detection
  using wav2vec 2.0 and data augmentation},''
\newblock in {\em Proc. Odyssey}, 2022, pp. 112--119.

\bibitem{kingma2014adam}
Diederik~P Kingma and Jimmy Ba,
\newblock ``{Adam: A method for stochastic optimization},''
\newblock in {\em Proc. ICLR}, 2014.

\bibitem{wang2021comparative}
Xin Wang and Junichi Yamagishi,
\newblock ``{A comparative study on recent neural spoofing countermeasures for
  synthetic speech detection},''
\newblock in {\em Proc. Interspeech}, 2021, pp. 4259--4263.

\bibitem{Tak2021}
Hemlata Tak, Madhu~R Kamble, Jose Patino, Massimiliano Todisco, and Nicholas
  W~D Evans,
\newblock ``{RawBoost: A Raw Data Boosting and Augmentation Method applied to
  Automatic Speaker Verification Anti-Spoofing},''
\newblock in {\em Proc. ICASSP}, 2022, pp. 6382--6386.

\bibitem{Wang2023}
Xin Wang and Junichi Yamagishi,
\newblock ``{Investigating Active-learning-based Training Data Selection for
  Speech Spoofing Countermeasure},''
\newblock in {\em Proc. SLT}, 2023, p. accepted.

\bibitem{pianese2022deepfake}
Alessandro Pianese, Davide Cozzolino, Giovanni Poggi, and Luisa Verdoliva,
\newblock ``{Deepfake audio detection by speaker verification},''
\newblock {\em arXiv preprint arXiv:2209.14098}, 2022.

\end{thebibliography}

\newpage
\clearpage
\appendix
\section{Appendix}
The appendix includes results of statistical analysis, additional results on the comparison of different vocoded training sets, and additional results on contrastive feature loss.

\subsection{Statistical Analysis on CM EER}

We followed the method in \citeapp{wang2021comparative} for the statistical analysis, but we directly analyzed the averaged EER over multiple runs. Intra-system analysis was not conducted. 
The significance test used a significance level $\alpha=0.05$ with Holm-Bonferroni correction\footnote{A tutorial notebook on the statistical analysis is available\url{https://github.com/nii-yamagishilab/project-NN-Pytorch-scripts/blob/master/tutorials/b2_anti_spoofing/chapter_a1_stats_test.ipynb}}.

Due to limited page size, analysis results based on the EERs in Table~\ref{tab:exp-1} are split into Table~\ref{tab:app-sig-1} and Table~\ref{tab:app-sig-2}. Each subtable corresponds to the analysis results on one test set.  Similarly, Tables~\ref{tab:app-sig-3} and \ref{tab:app-sig-4} plot the analysis results based on the EERs in Table~\ref{tab:exp-2}.

From Tables~\ref{tab:app-sig-1} to ~\ref{tab:app-sig-2}, we note that the differences caused by different training sets are in many cases statistically significant. 

\begin{table}[!h]
\centering
\setlength{\tabcolsep}{5pt}
\renewcommand{\arraystretch}{1.2}
\caption{Results of statistical significance analysis for EERs in Table~\ref{tab:exp-1} (Part 1). Grey block indicates statistical significant difference.}
%\resizebox{0.8\columnwidth}{!}
{
\begin{tabular}{ccccccc}
\toprule
 {\setELA}                &  \rotatebox{90}{\setTLAshort}  &  \rotatebox{90}{\setTWFshort}  &  \rotatebox{90}{\setTVIshort}  & \rotatebox{90}{\setTVIIIshort} & \rotatebox{90}{\setTVIVshort}  &  \rotatebox{90}{\setTVVshort} \\ 
\midrule
 \setTLAshort  & \cellcolor[rgb]{1.00, 1.00, 1.00}  & \cellcolor[rgb]{0.59, 0.59, 0.59}  & \cellcolor[rgb]{0.59, 0.59, 0.59}  & \cellcolor[rgb]{0.59, 0.59, 0.59}  & \cellcolor[rgb]{0.59, 0.59, 0.59}  & \cellcolor[rgb]{0.59, 0.59, 0.59} \\ 
 \setTWFshort  & \cellcolor[rgb]{0.59, 0.59, 0.59}  & \cellcolor[rgb]{1.00, 1.00, 1.00}  & \cellcolor[rgb]{0.59, 0.59, 0.59}  & \cellcolor[rgb]{0.59, 0.59, 0.59}  & \cellcolor[rgb]{0.59, 0.59, 0.59}  & \cellcolor[rgb]{0.59, 0.59, 0.59} \\ 
 \setTVIshort  & \cellcolor[rgb]{0.59, 0.59, 0.59}  & \cellcolor[rgb]{0.59, 0.59, 0.59}  & \cellcolor[rgb]{1.00, 1.00, 1.00}  & \cellcolor[rgb]{1.00, 1.00, 1.00}  & \cellcolor[rgb]{0.59, 0.59, 0.59}  & \cellcolor[rgb]{0.59, 0.59, 0.59} \\ 
\setTVIIIshort & \cellcolor[rgb]{0.59, 0.59, 0.59}  & \cellcolor[rgb]{0.59, 0.59, 0.59}  & \cellcolor[rgb]{1.00, 1.00, 1.00}  & \cellcolor[rgb]{1.00, 1.00, 1.00}  & \cellcolor[rgb]{0.59, 0.59, 0.59}  & \cellcolor[rgb]{0.59, 0.59, 0.59} \\ 
\setTVIVshort  & \cellcolor[rgb]{0.59, 0.59, 0.59}  & \cellcolor[rgb]{0.59, 0.59, 0.59}  & \cellcolor[rgb]{0.59, 0.59, 0.59}  & \cellcolor[rgb]{0.59, 0.59, 0.59}  & \cellcolor[rgb]{1.00, 1.00, 1.00}  & \cellcolor[rgb]{0.59, 0.59, 0.59} \\ 
 \setTVVshort  & \cellcolor[rgb]{0.59, 0.59, 0.59}  & \cellcolor[rgb]{0.59, 0.59, 0.59}  & \cellcolor[rgb]{0.59, 0.59, 0.59}  & \cellcolor[rgb]{0.59, 0.59, 0.59}  & \cellcolor[rgb]{0.59, 0.59, 0.59}  & \cellcolor[rgb]{1.00, 1.00, 1.00} \\ 
\bottomrule
\\
 {\setELAII}                             \\ 
\midrule
 \setTLAshort  & \cellcolor[rgb]{1.00, 1.00, 1.00}  & \cellcolor[rgb]{0.59, 0.59, 0.59}  & \cellcolor[rgb]{0.59, 0.59, 0.59}  & \cellcolor[rgb]{0.59, 0.59, 0.59}  & \cellcolor[rgb]{0.59, 0.59, 0.59}  & \cellcolor[rgb]{0.59, 0.59, 0.59} \\ 
 \setTWFshort  & \cellcolor[rgb]{0.59, 0.59, 0.59}  & \cellcolor[rgb]{1.00, 1.00, 1.00}  & \cellcolor[rgb]{0.59, 0.59, 0.59}  & \cellcolor[rgb]{0.59, 0.59, 0.59}  & \cellcolor[rgb]{0.59, 0.59, 0.59}  & \cellcolor[rgb]{0.59, 0.59, 0.59} \\ 
 \setTVIshort  & \cellcolor[rgb]{0.59, 0.59, 0.59}  & \cellcolor[rgb]{0.59, 0.59, 0.59}  & \cellcolor[rgb]{1.00, 1.00, 1.00}  & \cellcolor[rgb]{0.59, 0.59, 0.59}  & \cellcolor[rgb]{0.59, 0.59, 0.59}  & \cellcolor[rgb]{0.59, 0.59, 0.59} \\ 
\setTVIIIshort & \cellcolor[rgb]{0.59, 0.59, 0.59}  & \cellcolor[rgb]{0.59, 0.59, 0.59}  & \cellcolor[rgb]{0.59, 0.59, 0.59}  & \cellcolor[rgb]{1.00, 1.00, 1.00}  & \cellcolor[rgb]{0.59, 0.59, 0.59}  & \cellcolor[rgb]{0.59, 0.59, 0.59} \\ 
\setTVIVshort  & \cellcolor[rgb]{0.59, 0.59, 0.59}  & \cellcolor[rgb]{0.59, 0.59, 0.59}  & \cellcolor[rgb]{0.59, 0.59, 0.59}  & \cellcolor[rgb]{0.59, 0.59, 0.59}  & \cellcolor[rgb]{1.00, 1.00, 1.00}  & \cellcolor[rgb]{0.59, 0.59, 0.59} \\ 
 \setTVVshort  & \cellcolor[rgb]{0.59, 0.59, 0.59}  & \cellcolor[rgb]{0.59, 0.59, 0.59}  & \cellcolor[rgb]{0.59, 0.59, 0.59}  & \cellcolor[rgb]{0.59, 0.59, 0.59}  & \cellcolor[rgb]{0.59, 0.59, 0.59}  & \cellcolor[rgb]{1.00, 1.00, 1.00} \\ 
\bottomrule
\\
 {\setEDF}                             \\ 
\midrule
 \setTLAshort  & \cellcolor[rgb]{1.00, 1.00, 1.00}  & \cellcolor[rgb]{0.59, 0.59, 0.59}  & \cellcolor[rgb]{0.59, 0.59, 0.59}  & \cellcolor[rgb]{0.59, 0.59, 0.59}  & \cellcolor[rgb]{0.59, 0.59, 0.59}  & \cellcolor[rgb]{0.59, 0.59, 0.59} \\ 
 \setTWFshort  & \cellcolor[rgb]{0.59, 0.59, 0.59}  & \cellcolor[rgb]{1.00, 1.00, 1.00}  & \cellcolor[rgb]{0.59, 0.59, 0.59}  & \cellcolor[rgb]{0.59, 0.59, 0.59}  & \cellcolor[rgb]{0.59, 0.59, 0.59}  & \cellcolor[rgb]{0.59, 0.59, 0.59} \\ 
 \setTVIshort  & \cellcolor[rgb]{0.59, 0.59, 0.59}  & \cellcolor[rgb]{0.59, 0.59, 0.59}  & \cellcolor[rgb]{1.00, 1.00, 1.00}  & \cellcolor[rgb]{1.00, 1.00, 1.00}  & \cellcolor[rgb]{0.59, 0.59, 0.59}  & \cellcolor[rgb]{0.59, 0.59, 0.59} \\ 
\setTVIIIshort & \cellcolor[rgb]{0.59, 0.59, 0.59}  & \cellcolor[rgb]{0.59, 0.59, 0.59}  & \cellcolor[rgb]{1.00, 1.00, 1.00}  & \cellcolor[rgb]{1.00, 1.00, 1.00}  & \cellcolor[rgb]{0.59, 0.59, 0.59}  & \cellcolor[rgb]{0.59, 0.59, 0.59} \\ 
\setTVIVshort  & \cellcolor[rgb]{0.59, 0.59, 0.59}  & \cellcolor[rgb]{0.59, 0.59, 0.59}  & \cellcolor[rgb]{0.59, 0.59, 0.59}  & \cellcolor[rgb]{0.59, 0.59, 0.59}  & \cellcolor[rgb]{1.00, 1.00, 1.00}  & \cellcolor[rgb]{0.59, 0.59, 0.59} \\ 
 \setTVVshort  & \cellcolor[rgb]{0.59, 0.59, 0.59}  & \cellcolor[rgb]{0.59, 0.59, 0.59}  & \cellcolor[rgb]{0.59, 0.59, 0.59}  & \cellcolor[rgb]{0.59, 0.59, 0.59}  & \cellcolor[rgb]{0.59, 0.59, 0.59}  & \cellcolor[rgb]{1.00, 1.00, 1.00} \\ 
\bottomrule
\end{tabular}
\label{tab:app-sig-1}
}
\end{table}

\begin{table}[!t]
\centering
%\Large
\setlength{\tabcolsep}{5pt}
\renewcommand{\arraystretch}{1.2}
\caption{Results of statistical significance analysis for EERs in Table~\ref{tab:exp-1} (Part 2). Grey block indicates statistical significant difference.}
%\resizebox{0.8\columnwidth}{!}
{
\begin{tabular}{ccccccc}
\toprule
 {\setELATRIM}        &  \rotatebox{90}{\setTLAshort}  &  \rotatebox{90}{\setTWFshort}  &  \rotatebox{90}{\setTVIshort}  & \rotatebox{90}{\setTVIIIshort} & \rotatebox{90}{\setTVIVshort}  &  \rotatebox{90}{\setTVVshort} \\
\midrule
 \setTLAshort  & \cellcolor[rgb]{1.00, 1.00, 1.00}  & \cellcolor[rgb]{0.59, 0.59, 0.59}  & \cellcolor[rgb]{0.59, 0.59, 0.59}  & \cellcolor[rgb]{1.00, 1.00, 1.00}  & \cellcolor[rgb]{1.00, 1.00, 1.00}  & \cellcolor[rgb]{0.59, 0.59, 0.59} \\ 
 \setTWFshort  & \cellcolor[rgb]{0.59, 0.59, 0.59}  & \cellcolor[rgb]{1.00, 1.00, 1.00}  & \cellcolor[rgb]{0.59, 0.59, 0.59}  & \cellcolor[rgb]{0.59, 0.59, 0.59}  & \cellcolor[rgb]{0.59, 0.59, 0.59}  & \cellcolor[rgb]{0.59, 0.59, 0.59} \\ 
 \setTVIshort  & \cellcolor[rgb]{0.59, 0.59, 0.59}  & \cellcolor[rgb]{0.59, 0.59, 0.59}  & \cellcolor[rgb]{1.00, 1.00, 1.00}  & \cellcolor[rgb]{0.59, 0.59, 0.59}  & \cellcolor[rgb]{0.59, 0.59, 0.59}  & \cellcolor[rgb]{0.59, 0.59, 0.59} \\ 
\setTVIIIshort & \cellcolor[rgb]{1.00, 1.00, 1.00}  & \cellcolor[rgb]{0.59, 0.59, 0.59}  & \cellcolor[rgb]{0.59, 0.59, 0.59}  & \cellcolor[rgb]{1.00, 1.00, 1.00}  & \cellcolor[rgb]{0.59, 0.59, 0.59}  & \cellcolor[rgb]{0.59, 0.59, 0.59} \\ 
\setTVIVshort  & \cellcolor[rgb]{1.00, 1.00, 1.00}  & \cellcolor[rgb]{0.59, 0.59, 0.59}  & \cellcolor[rgb]{0.59, 0.59, 0.59}  & \cellcolor[rgb]{0.59, 0.59, 0.59}  & \cellcolor[rgb]{1.00, 1.00, 1.00}  & \cellcolor[rgb]{0.59, 0.59, 0.59} \\ 
 \setTVVshort  & \cellcolor[rgb]{0.59, 0.59, 0.59}  & \cellcolor[rgb]{0.59, 0.59, 0.59}  & \cellcolor[rgb]{0.59, 0.59, 0.59}  & \cellcolor[rgb]{0.59, 0.59, 0.59}  & \cellcolor[rgb]{0.59, 0.59, 0.59}  & \cellcolor[rgb]{1.00, 1.00, 1.00} \\ 
\bottomrule
\\
{\setELAHID} \\
\midrule
 \setTLAshort  & \cellcolor[rgb]{1.00, 1.00, 1.00}  & \cellcolor[rgb]{1.00, 1.00, 1.00}  & \cellcolor[rgb]{1.00, 1.00, 1.00}  & \cellcolor[rgb]{0.59, 0.59, 0.59}  & \cellcolor[rgb]{0.59, 0.59, 0.59}  & \cellcolor[rgb]{0.59, 0.59, 0.59} \\ 
 \setTWFshort  & \cellcolor[rgb]{1.00, 1.00, 1.00}  & \cellcolor[rgb]{1.00, 1.00, 1.00}  & \cellcolor[rgb]{1.00, 1.00, 1.00}  & \cellcolor[rgb]{0.59, 0.59, 0.59}  & \cellcolor[rgb]{0.59, 0.59, 0.59}  & \cellcolor[rgb]{0.59, 0.59, 0.59} \\ 
 \setTVIshort  & \cellcolor[rgb]{1.00, 1.00, 1.00}  & \cellcolor[rgb]{1.00, 1.00, 1.00}  & \cellcolor[rgb]{1.00, 1.00, 1.00}  & \cellcolor[rgb]{0.59, 0.59, 0.59}  & \cellcolor[rgb]{0.59, 0.59, 0.59}  & \cellcolor[rgb]{0.59, 0.59, 0.59} \\ 
\setTVIIIshort & \cellcolor[rgb]{0.59, 0.59, 0.59}  & \cellcolor[rgb]{0.59, 0.59, 0.59}  & \cellcolor[rgb]{0.59, 0.59, 0.59}  & \cellcolor[rgb]{1.00, 1.00, 1.00}  & \cellcolor[rgb]{1.00, 1.00, 1.00}  & \cellcolor[rgb]{1.00, 1.00, 1.00} \\ 
\setTVIVshort  & \cellcolor[rgb]{0.59, 0.59, 0.59}  & \cellcolor[rgb]{0.59, 0.59, 0.59}  & \cellcolor[rgb]{0.59, 0.59, 0.59}  & \cellcolor[rgb]{1.00, 1.00, 1.00}  & \cellcolor[rgb]{1.00, 1.00, 1.00}  & \cellcolor[rgb]{0.59, 0.59, 0.59} \\ 
 \setTVVshort  & \cellcolor[rgb]{0.59, 0.59, 0.59}  & \cellcolor[rgb]{0.59, 0.59, 0.59}  & \cellcolor[rgb]{0.59, 0.59, 0.59}  & \cellcolor[rgb]{1.00, 1.00, 1.00}  & \cellcolor[rgb]{0.59, 0.59, 0.59}  & \cellcolor[rgb]{1.00, 1.00, 1.00} \\ 
\bottomrule
\\
\setEDFHID \\
\midrule
 \setTLAshort  & \cellcolor[rgb]{1.00, 1.00, 1.00}  & \cellcolor[rgb]{0.59, 0.59, 0.59}  & \cellcolor[rgb]{1.00, 1.00, 1.00}  & \cellcolor[rgb]{0.59, 0.59, 0.59}  & \cellcolor[rgb]{0.59, 0.59, 0.59}  & \cellcolor[rgb]{0.59, 0.59, 0.59} \\ 
 \setTWFshort  & \cellcolor[rgb]{0.59, 0.59, 0.59}  & \cellcolor[rgb]{1.00, 1.00, 1.00}  & \cellcolor[rgb]{0.59, 0.59, 0.59}  & \cellcolor[rgb]{0.59, 0.59, 0.59}  & \cellcolor[rgb]{0.59, 0.59, 0.59}  & \cellcolor[rgb]{0.59, 0.59, 0.59} \\ 
 \setTVIshort  & \cellcolor[rgb]{1.00, 1.00, 1.00}  & \cellcolor[rgb]{0.59, 0.59, 0.59}  & \cellcolor[rgb]{1.00, 1.00, 1.00}  & \cellcolor[rgb]{0.59, 0.59, 0.59}  & \cellcolor[rgb]{0.59, 0.59, 0.59}  & \cellcolor[rgb]{0.59, 0.59, 0.59} \\ 
\setTVIIIshort & \cellcolor[rgb]{0.59, 0.59, 0.59}  & \cellcolor[rgb]{0.59, 0.59, 0.59}  & \cellcolor[rgb]{0.59, 0.59, 0.59}  & \cellcolor[rgb]{1.00, 1.00, 1.00}  & \cellcolor[rgb]{1.00, 1.00, 1.00}  & \cellcolor[rgb]{0.59, 0.59, 0.59} \\ 
\setTVIVshort  & \cellcolor[rgb]{0.59, 0.59, 0.59}  & \cellcolor[rgb]{0.59, 0.59, 0.59}  & \cellcolor[rgb]{0.59, 0.59, 0.59}  & \cellcolor[rgb]{1.00, 1.00, 1.00}  & \cellcolor[rgb]{1.00, 1.00, 1.00}  & \cellcolor[rgb]{0.59, 0.59, 0.59} \\ 
 \setTVVshort  & \cellcolor[rgb]{0.59, 0.59, 0.59}  & \cellcolor[rgb]{0.59, 0.59, 0.59}  & \cellcolor[rgb]{0.59, 0.59, 0.59}  & \cellcolor[rgb]{0.59, 0.59, 0.59}  & \cellcolor[rgb]{0.59, 0.59, 0.59}  & \cellcolor[rgb]{1.00, 1.00, 1.00} \\ 
\bottomrule
\\
\setEWFE \\
\midrule
 \setTLAshort  & \cellcolor[rgb]{1.00, 1.00, 1.00}  & \cellcolor[rgb]{0.59, 0.59, 0.59}  & \cellcolor[rgb]{0.59, 0.59, 0.59}  & \cellcolor[rgb]{0.59, 0.59, 0.59}  & \cellcolor[rgb]{0.59, 0.59, 0.59}  & \cellcolor[rgb]{0.59, 0.59, 0.59} \\ 
 \setTWFshort  & \cellcolor[rgb]{0.59, 0.59, 0.59}  & \cellcolor[rgb]{1.00, 1.00, 1.00}  & \cellcolor[rgb]{0.59, 0.59, 0.59}  & \cellcolor[rgb]{0.59, 0.59, 0.59}  & \cellcolor[rgb]{0.59, 0.59, 0.59}  & \cellcolor[rgb]{0.59, 0.59, 0.59} \\ 
 \setTVIshort  & \cellcolor[rgb]{0.59, 0.59, 0.59}  & \cellcolor[rgb]{0.59, 0.59, 0.59}  & \cellcolor[rgb]{1.00, 1.00, 1.00}  & \cellcolor[rgb]{0.59, 0.59, 0.59}  & \cellcolor[rgb]{0.59, 0.59, 0.59}  & \cellcolor[rgb]{0.59, 0.59, 0.59} \\ 
\setTVIIIshort & \cellcolor[rgb]{0.59, 0.59, 0.59}  & \cellcolor[rgb]{0.59, 0.59, 0.59}  & \cellcolor[rgb]{0.59, 0.59, 0.59}  & \cellcolor[rgb]{1.00, 1.00, 1.00}  & \cellcolor[rgb]{0.59, 0.59, 0.59}  & \cellcolor[rgb]{0.59, 0.59, 0.59} \\ 
\setTVIVshort  & \cellcolor[rgb]{0.59, 0.59, 0.59}  & \cellcolor[rgb]{0.59, 0.59, 0.59}  & \cellcolor[rgb]{0.59, 0.59, 0.59}  & \cellcolor[rgb]{0.59, 0.59, 0.59}  & \cellcolor[rgb]{1.00, 1.00, 1.00}  & \cellcolor[rgb]{0.59, 0.59, 0.59} \\ 
 \setTVVshort  & \cellcolor[rgb]{0.59, 0.59, 0.59}  & \cellcolor[rgb]{0.59, 0.59, 0.59}  & \cellcolor[rgb]{0.59, 0.59, 0.59}  & \cellcolor[rgb]{0.59, 0.59, 0.59}  & \cellcolor[rgb]{0.59, 0.59, 0.59}  & \cellcolor[rgb]{1.00, 1.00, 1.00} \\ 
\bottomrule
\\
\setEDFE \\
\midrule
 \setTLAshort  & \cellcolor[rgb]{1.00, 1.00, 1.00}  & \cellcolor[rgb]{0.59, 0.59, 0.59}  & \cellcolor[rgb]{0.59, 0.59, 0.59}  & \cellcolor[rgb]{0.59, 0.59, 0.59}  & \cellcolor[rgb]{0.59, 0.59, 0.59}  & \cellcolor[rgb]{0.59, 0.59, 0.59} \\ 
 \setTWFshort  & \cellcolor[rgb]{0.59, 0.59, 0.59}  & \cellcolor[rgb]{1.00, 1.00, 1.00}  & \cellcolor[rgb]{0.59, 0.59, 0.59}  & \cellcolor[rgb]{0.59, 0.59, 0.59}  & \cellcolor[rgb]{0.59, 0.59, 0.59}  & \cellcolor[rgb]{1.00, 1.00, 1.00} \\ 
 \setTVIshort  & \cellcolor[rgb]{0.59, 0.59, 0.59}  & \cellcolor[rgb]{0.59, 0.59, 0.59}  & \cellcolor[rgb]{1.00, 1.00, 1.00}  & \cellcolor[rgb]{0.59, 0.59, 0.59}  & \cellcolor[rgb]{0.59, 0.59, 0.59}  & \cellcolor[rgb]{0.59, 0.59, 0.59} \\ 
\setTVIIIshort & \cellcolor[rgb]{0.59, 0.59, 0.59}  & \cellcolor[rgb]{0.59, 0.59, 0.59}  & \cellcolor[rgb]{0.59, 0.59, 0.59}  & \cellcolor[rgb]{1.00, 1.00, 1.00}  & \cellcolor[rgb]{0.59, 0.59, 0.59}  & \cellcolor[rgb]{0.59, 0.59, 0.59} \\ 
\setTVIVshort  & \cellcolor[rgb]{0.59, 0.59, 0.59}  & \cellcolor[rgb]{0.59, 0.59, 0.59}  & \cellcolor[rgb]{0.59, 0.59, 0.59}  & \cellcolor[rgb]{0.59, 0.59, 0.59}  & \cellcolor[rgb]{1.00, 1.00, 1.00}  & \cellcolor[rgb]{0.59, 0.59, 0.59} \\ 
 \setTVVshort  & \cellcolor[rgb]{0.59, 0.59, 0.59}  & \cellcolor[rgb]{1.00, 1.00, 1.00}  & \cellcolor[rgb]{0.59, 0.59, 0.59}  & \cellcolor[rgb]{0.59, 0.59, 0.59}  & \cellcolor[rgb]{0.59, 0.59, 0.59}  & \cellcolor[rgb]{1.00, 1.00, 1.00} \\ 
\bottomrule
\end{tabular}
}
\label{tab:app-sig-2}
\end{table}

\newpage
\clearpage

\begin{table}[!t]
\centering
\caption{Results of statistical significance analysis for EERs in Table~\ref{tab:exp-2} (Part 1). Grey block indicates statistical significant difference.}
\setlength{\tabcolsep}{1pt}
\renewcommand{\arraystretch}{1.2}
\resizebox{0.90\columnwidth}{!}
{
\begin{tabular}{cccccccc}
\toprule
\multicolumn{1}{l}{Training criterion}  & \multicolumn{4}{c}{$\mathcal{L}_{\text{CE}}$ }   & \multicolumn{3}{c}{$\mathcal{L}_{\text{CE}} + \mathcal{L}_{\text{CF}}$ }   \\
  \cmidrule(lr){2-5}\cmidrule(lr){6-8} 
  \multicolumn{1}{l}{Data augmentation} &  \multicolumn{2}{c}{$\times$} & \multicolumn{2}{c}{RawBoost} &  \multicolumn{3}{c}{RawBoost}  \\
  \cmidrule(lr){2-3}\cmidrule(lr){4-8} 
   \multicolumn{1}{l}{Training set}     & \setTLA    & \setTVV & \setTLA & \setTVV & \setTLA & \setTVV & \setTVV\\ 
 \multicolumn{1}{l}{Bona-spoof paired}   &  $\times$  &  $\times$  &  $\times$  &  $\times$  &  $\times$  &  $\times$ &  $\checkmark$ \\
\midrule
 {\setELA}      & \selfcircle{1} & \selfcircle{2} & \selfcircle{3} & \selfcircle{4} & \selfcircle{5} & \selfcircle{6} & \selfcircle{7}\\ 
\midrule
\selfcircle{1} & \cellcolor[rgb]{1.00, 1.00, 1.00}  & \cellcolor[rgb]{0.59, 0.59, 0.59}  & \cellcolor[rgb]{0.59, 0.59, 0.59}  & \cellcolor[rgb]{0.59, 0.59, 0.59}  & \cellcolor[rgb]{0.59, 0.59, 0.59}  & \cellcolor[rgb]{1.00, 1.00, 1.00}  & \cellcolor[rgb]{0.59, 0.59, 0.59} \\ 
\selfcircle{2} & \cellcolor[rgb]{0.59, 0.59, 0.59}  & \cellcolor[rgb]{1.00, 1.00, 1.00}  & \cellcolor[rgb]{0.59, 0.59, 0.59}  & \cellcolor[rgb]{0.59, 0.59, 0.59}  & \cellcolor[rgb]{0.59, 0.59, 0.59}  & \cellcolor[rgb]{0.59, 0.59, 0.59}  & \cellcolor[rgb]{0.59, 0.59, 0.59} \\ 
\selfcircle{3} & \cellcolor[rgb]{0.59, 0.59, 0.59}  & \cellcolor[rgb]{0.59, 0.59, 0.59}  & \cellcolor[rgb]{1.00, 1.00, 1.00}  & \cellcolor[rgb]{0.59, 0.59, 0.59}  & \cellcolor[rgb]{1.00, 1.00, 1.00}  & \cellcolor[rgb]{0.59, 0.59, 0.59}  & \cellcolor[rgb]{0.59, 0.59, 0.59} \\ 
\selfcircle{4} & \cellcolor[rgb]{0.59, 0.59, 0.59}  & \cellcolor[rgb]{0.59, 0.59, 0.59}  & \cellcolor[rgb]{0.59, 0.59, 0.59}  & \cellcolor[rgb]{1.00, 1.00, 1.00}  & \cellcolor[rgb]{0.59, 0.59, 0.59}  & \cellcolor[rgb]{0.59, 0.59, 0.59}  & \cellcolor[rgb]{0.59, 0.59, 0.59} \\ 
\selfcircle{5} & \cellcolor[rgb]{0.59, 0.59, 0.59}  & \cellcolor[rgb]{0.59, 0.59, 0.59}  & \cellcolor[rgb]{1.00, 1.00, 1.00}  & \cellcolor[rgb]{0.59, 0.59, 0.59}  & \cellcolor[rgb]{1.00, 1.00, 1.00}  & \cellcolor[rgb]{0.59, 0.59, 0.59}  & \cellcolor[rgb]{0.59, 0.59, 0.59} \\ 
\selfcircle{6} & \cellcolor[rgb]{1.00, 1.00, 1.00}  & \cellcolor[rgb]{0.59, 0.59, 0.59}  & \cellcolor[rgb]{0.59, 0.59, 0.59}  & \cellcolor[rgb]{0.59, 0.59, 0.59}  & \cellcolor[rgb]{0.59, 0.59, 0.59}  & \cellcolor[rgb]{1.00, 1.00, 1.00}  & \cellcolor[rgb]{0.59, 0.59, 0.59} \\ 
\selfcircle{7} & \cellcolor[rgb]{0.59, 0.59, 0.59}  & \cellcolor[rgb]{0.59, 0.59, 0.59}  & \cellcolor[rgb]{0.59, 0.59, 0.59}  & \cellcolor[rgb]{0.59, 0.59, 0.59}  & \cellcolor[rgb]{0.59, 0.59, 0.59}  & \cellcolor[rgb]{0.59, 0.59, 0.59}  & \cellcolor[rgb]{1.00, 1.00, 1.00} \\ 
\bottomrule
\\
 {\setELAII}                     \\
\midrule
\selfcircle{1} & \cellcolor[rgb]{1.00, 1.00, 1.00}  & \cellcolor[rgb]{0.59, 0.59, 0.59}  & \cellcolor[rgb]{0.59, 0.59, 0.59}  & \cellcolor[rgb]{0.59, 0.59, 0.59}  & \cellcolor[rgb]{0.59, 0.59, 0.59}  & \cellcolor[rgb]{0.59, 0.59, 0.59}  & \cellcolor[rgb]{0.59, 0.59, 0.59} \\ 
\selfcircle{2} & \cellcolor[rgb]{0.59, 0.59, 0.59}  & \cellcolor[rgb]{1.00, 1.00, 1.00}  & \cellcolor[rgb]{0.59, 0.59, 0.59}  & \cellcolor[rgb]{0.59, 0.59, 0.59}  & \cellcolor[rgb]{0.59, 0.59, 0.59}  & \cellcolor[rgb]{0.59, 0.59, 0.59}  & \cellcolor[rgb]{0.59, 0.59, 0.59} \\ 
\selfcircle{3} & \cellcolor[rgb]{0.59, 0.59, 0.59}  & \cellcolor[rgb]{0.59, 0.59, 0.59}  & \cellcolor[rgb]{1.00, 1.00, 1.00}  & \cellcolor[rgb]{0.59, 0.59, 0.59}  & \cellcolor[rgb]{0.59, 0.59, 0.59}  & \cellcolor[rgb]{0.59, 0.59, 0.59}  & \cellcolor[rgb]{0.59, 0.59, 0.59} \\ 
\selfcircle{4} & \cellcolor[rgb]{0.59, 0.59, 0.59}  & \cellcolor[rgb]{0.59, 0.59, 0.59}  & \cellcolor[rgb]{0.59, 0.59, 0.59}  & \cellcolor[rgb]{1.00, 1.00, 1.00}  & \cellcolor[rgb]{0.59, 0.59, 0.59}  & \cellcolor[rgb]{1.00, 1.00, 1.00}  & \cellcolor[rgb]{0.59, 0.59, 0.59} \\ 
\selfcircle{5} & \cellcolor[rgb]{0.59, 0.59, 0.59}  & \cellcolor[rgb]{0.59, 0.59, 0.59}  & \cellcolor[rgb]{0.59, 0.59, 0.59}  & \cellcolor[rgb]{0.59, 0.59, 0.59}  & \cellcolor[rgb]{1.00, 1.00, 1.00}  & \cellcolor[rgb]{0.59, 0.59, 0.59}  & \cellcolor[rgb]{0.59, 0.59, 0.59} \\ 
\selfcircle{6} & \cellcolor[rgb]{0.59, 0.59, 0.59}  & \cellcolor[rgb]{0.59, 0.59, 0.59}  & \cellcolor[rgb]{0.59, 0.59, 0.59}  & \cellcolor[rgb]{1.00, 1.00, 1.00}  & \cellcolor[rgb]{0.59, 0.59, 0.59}  & \cellcolor[rgb]{1.00, 1.00, 1.00}  & \cellcolor[rgb]{0.59, 0.59, 0.59} \\ 
\selfcircle{7} & \cellcolor[rgb]{0.59, 0.59, 0.59}  & \cellcolor[rgb]{0.59, 0.59, 0.59}  & \cellcolor[rgb]{0.59, 0.59, 0.59}  & \cellcolor[rgb]{0.59, 0.59, 0.59}  & \cellcolor[rgb]{0.59, 0.59, 0.59}  & \cellcolor[rgb]{0.59, 0.59, 0.59}  & \cellcolor[rgb]{1.00, 1.00, 1.00} \\ 
\bottomrule
\\
{\setEDF}                \\
\midrule
\selfcircle{1} & \cellcolor[rgb]{1.00, 1.00, 1.00}  & \cellcolor[rgb]{0.59, 0.59, 0.59}  & \cellcolor[rgb]{0.59, 0.59, 0.59}  & \cellcolor[rgb]{0.59, 0.59, 0.59}  & \cellcolor[rgb]{0.59, 0.59, 0.59}  & \cellcolor[rgb]{1.00, 1.00, 1.00}  & \cellcolor[rgb]{0.59, 0.59, 0.59} \\ 
\selfcircle{2} & \cellcolor[rgb]{0.59, 0.59, 0.59}  & \cellcolor[rgb]{1.00, 1.00, 1.00}  & \cellcolor[rgb]{0.59, 0.59, 0.59}  & \cellcolor[rgb]{0.59, 0.59, 0.59}  & \cellcolor[rgb]{0.59, 0.59, 0.59}  & \cellcolor[rgb]{0.59, 0.59, 0.59}  & \cellcolor[rgb]{0.59, 0.59, 0.59} \\ 
\selfcircle{3} & \cellcolor[rgb]{0.59, 0.59, 0.59}  & \cellcolor[rgb]{0.59, 0.59, 0.59}  & \cellcolor[rgb]{1.00, 1.00, 1.00}  & \cellcolor[rgb]{0.59, 0.59, 0.59}  & \cellcolor[rgb]{0.59, 0.59, 0.59}  & \cellcolor[rgb]{0.59, 0.59, 0.59}  & \cellcolor[rgb]{0.59, 0.59, 0.59} \\ 
\selfcircle{4} & \cellcolor[rgb]{0.59, 0.59, 0.59}  & \cellcolor[rgb]{0.59, 0.59, 0.59}  & \cellcolor[rgb]{0.59, 0.59, 0.59}  & \cellcolor[rgb]{1.00, 1.00, 1.00}  & \cellcolor[rgb]{0.59, 0.59, 0.59}  & \cellcolor[rgb]{0.59, 0.59, 0.59}  & \cellcolor[rgb]{0.59, 0.59, 0.59} \\ 
\selfcircle{5} & \cellcolor[rgb]{0.59, 0.59, 0.59}  & \cellcolor[rgb]{0.59, 0.59, 0.59}  & \cellcolor[rgb]{0.59, 0.59, 0.59}  & \cellcolor[rgb]{0.59, 0.59, 0.59}  & \cellcolor[rgb]{1.00, 1.00, 1.00}  & \cellcolor[rgb]{0.59, 0.59, 0.59}  & \cellcolor[rgb]{0.59, 0.59, 0.59} \\ 
\selfcircle{6} & \cellcolor[rgb]{1.00, 1.00, 1.00}  & \cellcolor[rgb]{0.59, 0.59, 0.59}  & \cellcolor[rgb]{0.59, 0.59, 0.59}  & \cellcolor[rgb]{0.59, 0.59, 0.59}  & \cellcolor[rgb]{0.59, 0.59, 0.59}  & \cellcolor[rgb]{1.00, 1.00, 1.00}  & \cellcolor[rgb]{0.59, 0.59, 0.59} \\ 
\selfcircle{7} & \cellcolor[rgb]{0.59, 0.59, 0.59}  & \cellcolor[rgb]{0.59, 0.59, 0.59}  & \cellcolor[rgb]{0.59, 0.59, 0.59}  & \cellcolor[rgb]{0.59, 0.59, 0.59}  & \cellcolor[rgb]{0.59, 0.59, 0.59}  & \cellcolor[rgb]{0.59, 0.59, 0.59}  & \cellcolor[rgb]{1.00, 1.00, 1.00} \\ 
\bottomrule
\end{tabular}
}
\label{tab:app-sig-3}
\end{table}

\begin{table}[!t]
\centering
\caption{Results of statistical significance analysis for EERs in Table~\ref{tab:exp-2} (Part 1). Grey block indicates statistical significant difference.}
\setlength{\tabcolsep}{1pt}
\renewcommand{\arraystretch}{1.2}
\resizebox{0.90\columnwidth}{!}
{
\begin{tabular}{cccccccc}
\toprule
\multicolumn{1}{l}{Training criterion}  & \multicolumn{4}{c}{$\mathcal{L}_{\text{CE}}$ }   & \multicolumn{3}{c}{$\mathcal{L}_{\text{CE}} + \mathcal{L}_{\text{CF}}$ }   \\
  \cmidrule(lr){2-5}\cmidrule(lr){6-8} 
  \multicolumn{1}{l}{Data augmentation} &  \multicolumn{2}{c}{$\times$} & \multicolumn{2}{c}{RawBoost} &  \multicolumn{3}{c}{RawBoost}  \\
  \cmidrule(lr){2-3}\cmidrule(lr){4-8} 
   \multicolumn{1}{l}{Training set}     & \setTLA    & \setTVV & \setTLA & \setTVV & \setTLA & \setTVV & \setTVV\\ 
 \multicolumn{1}{l}{Bona-spoof paired}   &  $\times$  &  $\times$  &  $\times$  &  $\times$  &  $\times$  &  $\times$ &  $\checkmark$ \\
\midrule
 {\setELATRIM}    & \selfcircle{1} & \selfcircle{2} & \selfcircle{3} & \selfcircle{4} & \selfcircle{5} & \selfcircle{6} & \selfcircle{7}\\ 
\midrule
\selfcircle{1} & \cellcolor[rgb]{1.00, 1.00, 1.00}  & \cellcolor[rgb]{0.59, 0.59, 0.59}  & \cellcolor[rgb]{0.59, 0.59, 0.59}  & \cellcolor[rgb]{0.59, 0.59, 0.59}  & \cellcolor[rgb]{0.59, 0.59, 0.59}  & \cellcolor[rgb]{0.59, 0.59, 0.59}  & \cellcolor[rgb]{0.59, 0.59, 0.59} \\ 
\selfcircle{2} & \cellcolor[rgb]{0.59, 0.59, 0.59}  & \cellcolor[rgb]{1.00, 1.00, 1.00}  & \cellcolor[rgb]{1.00, 1.00, 1.00}  & \cellcolor[rgb]{0.59, 0.59, 0.59}  & \cellcolor[rgb]{1.00, 1.00, 1.00}  & \cellcolor[rgb]{0.59, 0.59, 0.59}  & \cellcolor[rgb]{0.59, 0.59, 0.59} \\ 
\selfcircle{3} & \cellcolor[rgb]{0.59, 0.59, 0.59}  & \cellcolor[rgb]{1.00, 1.00, 1.00}  & \cellcolor[rgb]{1.00, 1.00, 1.00}  & \cellcolor[rgb]{0.59, 0.59, 0.59}  & \cellcolor[rgb]{1.00, 1.00, 1.00}  & \cellcolor[rgb]{0.59, 0.59, 0.59}  & \cellcolor[rgb]{0.59, 0.59, 0.59} \\ 
\selfcircle{4} & \cellcolor[rgb]{0.59, 0.59, 0.59}  & \cellcolor[rgb]{0.59, 0.59, 0.59}  & \cellcolor[rgb]{0.59, 0.59, 0.59}  & \cellcolor[rgb]{1.00, 1.00, 1.00}  & \cellcolor[rgb]{0.59, 0.59, 0.59}  & \cellcolor[rgb]{0.59, 0.59, 0.59}  & \cellcolor[rgb]{0.59, 0.59, 0.59} \\ 
\selfcircle{5} & \cellcolor[rgb]{0.59, 0.59, 0.59}  & \cellcolor[rgb]{1.00, 1.00, 1.00}  & \cellcolor[rgb]{1.00, 1.00, 1.00}  & \cellcolor[rgb]{0.59, 0.59, 0.59}  & \cellcolor[rgb]{1.00, 1.00, 1.00}  & \cellcolor[rgb]{0.59, 0.59, 0.59}  & \cellcolor[rgb]{0.59, 0.59, 0.59} \\ 
\selfcircle{6} & \cellcolor[rgb]{0.59, 0.59, 0.59}  & \cellcolor[rgb]{0.59, 0.59, 0.59}  & \cellcolor[rgb]{0.59, 0.59, 0.59}  & \cellcolor[rgb]{0.59, 0.59, 0.59}  & \cellcolor[rgb]{0.59, 0.59, 0.59}  & \cellcolor[rgb]{1.00, 1.00, 1.00}  & \cellcolor[rgb]{0.59, 0.59, 0.59} \\ 
\selfcircle{7} & \cellcolor[rgb]{0.59, 0.59, 0.59}  & \cellcolor[rgb]{0.59, 0.59, 0.59}  & \cellcolor[rgb]{0.59, 0.59, 0.59}  & \cellcolor[rgb]{0.59, 0.59, 0.59}  & \cellcolor[rgb]{0.59, 0.59, 0.59}  & \cellcolor[rgb]{0.59, 0.59, 0.59}  & \cellcolor[rgb]{1.00, 1.00, 1.00} \\ 
\bottomrule
\\
{\setELAHID} \\
\midrule
\selfcircle{1} & \cellcolor[rgb]{1.00, 1.00, 1.00}  & \cellcolor[rgb]{0.59, 0.59, 0.59}  & \cellcolor[rgb]{0.59, 0.59, 0.59}  & \cellcolor[rgb]{0.59, 0.59, 0.59}  & \cellcolor[rgb]{1.00, 1.00, 1.00}  & \cellcolor[rgb]{0.59, 0.59, 0.59}  & \cellcolor[rgb]{0.59, 0.59, 0.59} \\ 
\selfcircle{2} & \cellcolor[rgb]{0.59, 0.59, 0.59}  & \cellcolor[rgb]{1.00, 1.00, 1.00}  & \cellcolor[rgb]{1.00, 1.00, 1.00}  & \cellcolor[rgb]{0.59, 0.59, 0.59}  & \cellcolor[rgb]{0.59, 0.59, 0.59}  & \cellcolor[rgb]{0.59, 0.59, 0.59}  & \cellcolor[rgb]{0.59, 0.59, 0.59} \\ 
\selfcircle{3} & \cellcolor[rgb]{0.59, 0.59, 0.59}  & \cellcolor[rgb]{1.00, 1.00, 1.00}  & \cellcolor[rgb]{1.00, 1.00, 1.00}  & \cellcolor[rgb]{1.00, 1.00, 1.00}  & \cellcolor[rgb]{0.59, 0.59, 0.59}  & \cellcolor[rgb]{0.59, 0.59, 0.59}  & \cellcolor[rgb]{0.59, 0.59, 0.59} \\ 
\selfcircle{4} & \cellcolor[rgb]{0.59, 0.59, 0.59}  & \cellcolor[rgb]{0.59, 0.59, 0.59}  & \cellcolor[rgb]{1.00, 1.00, 1.00}  & \cellcolor[rgb]{1.00, 1.00, 1.00}  & \cellcolor[rgb]{0.59, 0.59, 0.59}  & \cellcolor[rgb]{0.59, 0.59, 0.59}  & \cellcolor[rgb]{0.59, 0.59, 0.59} \\ 
\selfcircle{5} & \cellcolor[rgb]{1.00, 1.00, 1.00}  & \cellcolor[rgb]{0.59, 0.59, 0.59}  & \cellcolor[rgb]{0.59, 0.59, 0.59}  & \cellcolor[rgb]{0.59, 0.59, 0.59}  & \cellcolor[rgb]{1.00, 1.00, 1.00}  & \cellcolor[rgb]{0.59, 0.59, 0.59}  & \cellcolor[rgb]{0.59, 0.59, 0.59} \\ 
\selfcircle{6} & \cellcolor[rgb]{0.59, 0.59, 0.59}  & \cellcolor[rgb]{0.59, 0.59, 0.59}  & \cellcolor[rgb]{0.59, 0.59, 0.59}  & \cellcolor[rgb]{0.59, 0.59, 0.59}  & \cellcolor[rgb]{0.59, 0.59, 0.59}  & \cellcolor[rgb]{1.00, 1.00, 1.00}  & \cellcolor[rgb]{1.00, 1.00, 1.00} \\ 
\selfcircle{7} & \cellcolor[rgb]{0.59, 0.59, 0.59}  & \cellcolor[rgb]{0.59, 0.59, 0.59}  & \cellcolor[rgb]{0.59, 0.59, 0.59}  & \cellcolor[rgb]{0.59, 0.59, 0.59}  & \cellcolor[rgb]{0.59, 0.59, 0.59}  & \cellcolor[rgb]{1.00, 1.00, 1.00}  & \cellcolor[rgb]{1.00, 1.00, 1.00} \\ 
\bottomrule
\\
\setEDFHID \\
\midrule
\selfcircle{1} & \cellcolor[rgb]{1.00, 1.00, 1.00}  & \cellcolor[rgb]{0.59, 0.59, 0.59}  & \cellcolor[rgb]{0.59, 0.59, 0.59}  & \cellcolor[rgb]{0.59, 0.59, 0.59}  & \cellcolor[rgb]{0.59, 0.59, 0.59}  & \cellcolor[rgb]{0.59, 0.59, 0.59}  & \cellcolor[rgb]{0.59, 0.59, 0.59} \\ 
\selfcircle{2} & \cellcolor[rgb]{0.59, 0.59, 0.59}  & \cellcolor[rgb]{1.00, 1.00, 1.00}  & \cellcolor[rgb]{0.59, 0.59, 0.59}  & \cellcolor[rgb]{0.59, 0.59, 0.59}  & \cellcolor[rgb]{1.00, 1.00, 1.00}  & \cellcolor[rgb]{0.59, 0.59, 0.59}  & \cellcolor[rgb]{0.59, 0.59, 0.59} \\ 
\selfcircle{3} & \cellcolor[rgb]{0.59, 0.59, 0.59}  & \cellcolor[rgb]{0.59, 0.59, 0.59}  & \cellcolor[rgb]{1.00, 1.00, 1.00}  & \cellcolor[rgb]{1.00, 1.00, 1.00}  & \cellcolor[rgb]{0.59, 0.59, 0.59}  & \cellcolor[rgb]{0.59, 0.59, 0.59}  & \cellcolor[rgb]{0.59, 0.59, 0.59} \\ 
\selfcircle{4} & \cellcolor[rgb]{0.59, 0.59, 0.59}  & \cellcolor[rgb]{0.59, 0.59, 0.59}  & \cellcolor[rgb]{1.00, 1.00, 1.00}  & \cellcolor[rgb]{1.00, 1.00, 1.00}  & \cellcolor[rgb]{0.59, 0.59, 0.59}  & \cellcolor[rgb]{0.59, 0.59, 0.59}  & \cellcolor[rgb]{0.59, 0.59, 0.59} \\ 
\selfcircle{5} & \cellcolor[rgb]{0.59, 0.59, 0.59}  & \cellcolor[rgb]{1.00, 1.00, 1.00}  & \cellcolor[rgb]{0.59, 0.59, 0.59}  & \cellcolor[rgb]{0.59, 0.59, 0.59}  & \cellcolor[rgb]{1.00, 1.00, 1.00}  & \cellcolor[rgb]{0.59, 0.59, 0.59}  & \cellcolor[rgb]{0.59, 0.59, 0.59} \\ 
\selfcircle{6} & \cellcolor[rgb]{0.59, 0.59, 0.59}  & \cellcolor[rgb]{0.59, 0.59, 0.59}  & \cellcolor[rgb]{0.59, 0.59, 0.59}  & \cellcolor[rgb]{0.59, 0.59, 0.59}  & \cellcolor[rgb]{0.59, 0.59, 0.59}  & \cellcolor[rgb]{1.00, 1.00, 1.00}  & \cellcolor[rgb]{1.00, 1.00, 1.00} \\ 
\selfcircle{7} & \cellcolor[rgb]{0.59, 0.59, 0.59}  & \cellcolor[rgb]{0.59, 0.59, 0.59}  & \cellcolor[rgb]{0.59, 0.59, 0.59}  & \cellcolor[rgb]{0.59, 0.59, 0.59}  & \cellcolor[rgb]{0.59, 0.59, 0.59}  & \cellcolor[rgb]{1.00, 1.00, 1.00}  & \cellcolor[rgb]{1.00, 1.00, 1.00} \\ 
\bottomrule
\\
\setEWFE \\
\midrule
\selfcircle{1} & \cellcolor[rgb]{1.00, 1.00, 1.00}  & \cellcolor[rgb]{0.59, 0.59, 0.59}  & \cellcolor[rgb]{0.59, 0.59, 0.59}  & \cellcolor[rgb]{0.59, 0.59, 0.59}  & \cellcolor[rgb]{0.59, 0.59, 0.59}  & \cellcolor[rgb]{0.59, 0.59, 0.59}  & \cellcolor[rgb]{0.59, 0.59, 0.59} \\ 
\selfcircle{2} & \cellcolor[rgb]{0.59, 0.59, 0.59}  & \cellcolor[rgb]{1.00, 1.00, 1.00}  & \cellcolor[rgb]{0.59, 0.59, 0.59}  & \cellcolor[rgb]{0.59, 0.59, 0.59}  & \cellcolor[rgb]{0.59, 0.59, 0.59}  & \cellcolor[rgb]{0.59, 0.59, 0.59}  & \cellcolor[rgb]{0.59, 0.59, 0.59} \\ 
\selfcircle{3} & \cellcolor[rgb]{0.59, 0.59, 0.59}  & \cellcolor[rgb]{0.59, 0.59, 0.59}  & \cellcolor[rgb]{1.00, 1.00, 1.00}  & \cellcolor[rgb]{0.59, 0.59, 0.59}  & \cellcolor[rgb]{0.59, 0.59, 0.59}  & \cellcolor[rgb]{0.59, 0.59, 0.59}  & \cellcolor[rgb]{0.59, 0.59, 0.59} \\ 
\selfcircle{4} & \cellcolor[rgb]{0.59, 0.59, 0.59}  & \cellcolor[rgb]{0.59, 0.59, 0.59}  & \cellcolor[rgb]{0.59, 0.59, 0.59}  & \cellcolor[rgb]{1.00, 1.00, 1.00}  & \cellcolor[rgb]{0.59, 0.59, 0.59}  & \cellcolor[rgb]{0.59, 0.59, 0.59}  & \cellcolor[rgb]{0.59, 0.59, 0.59} \\ 
\selfcircle{5} & \cellcolor[rgb]{0.59, 0.59, 0.59}  & \cellcolor[rgb]{0.59, 0.59, 0.59}  & \cellcolor[rgb]{0.59, 0.59, 0.59}  & \cellcolor[rgb]{0.59, 0.59, 0.59}  & \cellcolor[rgb]{1.00, 1.00, 1.00}  & \cellcolor[rgb]{0.59, 0.59, 0.59}  & \cellcolor[rgb]{0.59, 0.59, 0.59} \\ 
\selfcircle{6} & \cellcolor[rgb]{0.59, 0.59, 0.59}  & \cellcolor[rgb]{0.59, 0.59, 0.59}  & \cellcolor[rgb]{0.59, 0.59, 0.59}  & \cellcolor[rgb]{0.59, 0.59, 0.59}  & \cellcolor[rgb]{0.59, 0.59, 0.59}  & \cellcolor[rgb]{1.00, 1.00, 1.00}  & \cellcolor[rgb]{0.59, 0.59, 0.59} \\ 
\selfcircle{7} & \cellcolor[rgb]{0.59, 0.59, 0.59}  & \cellcolor[rgb]{0.59, 0.59, 0.59}  & \cellcolor[rgb]{0.59, 0.59, 0.59}  & \cellcolor[rgb]{0.59, 0.59, 0.59}  & \cellcolor[rgb]{0.59, 0.59, 0.59}  & \cellcolor[rgb]{0.59, 0.59, 0.59}  & \cellcolor[rgb]{1.00, 1.00, 1.00} \\ 
\bottomrule
\\
\setEDFE \\
\midrule
\selfcircle{1} & \cellcolor[rgb]{1.00, 1.00, 1.00}  & \cellcolor[rgb]{0.59, 0.59, 0.59}  & \cellcolor[rgb]{0.59, 0.59, 0.59}  & \cellcolor[rgb]{0.59, 0.59, 0.59}  & \cellcolor[rgb]{0.59, 0.59, 0.59}  & \cellcolor[rgb]{0.59, 0.59, 0.59}  & \cellcolor[rgb]{0.59, 0.59, 0.59} \\ 
\selfcircle{2} & \cellcolor[rgb]{0.59, 0.59, 0.59}  & \cellcolor[rgb]{1.00, 1.00, 1.00}  & \cellcolor[rgb]{0.59, 0.59, 0.59}  & \cellcolor[rgb]{0.59, 0.59, 0.59}  & \cellcolor[rgb]{0.59, 0.59, 0.59}  & \cellcolor[rgb]{0.59, 0.59, 0.59}  & \cellcolor[rgb]{0.59, 0.59, 0.59} \\ 
\selfcircle{3} & \cellcolor[rgb]{0.59, 0.59, 0.59}  & \cellcolor[rgb]{0.59, 0.59, 0.59}  & \cellcolor[rgb]{1.00, 1.00, 1.00}  & \cellcolor[rgb]{0.59, 0.59, 0.59}  & \cellcolor[rgb]{0.59, 0.59, 0.59}  & \cellcolor[rgb]{0.59, 0.59, 0.59}  & \cellcolor[rgb]{0.59, 0.59, 0.59} \\ 
\selfcircle{4} & \cellcolor[rgb]{0.59, 0.59, 0.59}  & \cellcolor[rgb]{0.59, 0.59, 0.59}  & \cellcolor[rgb]{0.59, 0.59, 0.59}  & \cellcolor[rgb]{1.00, 1.00, 1.00}  & \cellcolor[rgb]{0.59, 0.59, 0.59}  & \cellcolor[rgb]{0.59, 0.59, 0.59}  & \cellcolor[rgb]{0.59, 0.59, 0.59} \\ 
\selfcircle{5} & \cellcolor[rgb]{0.59, 0.59, 0.59}  & \cellcolor[rgb]{0.59, 0.59, 0.59}  & \cellcolor[rgb]{0.59, 0.59, 0.59}  & \cellcolor[rgb]{0.59, 0.59, 0.59}  & \cellcolor[rgb]{1.00, 1.00, 1.00}  & \cellcolor[rgb]{0.59, 0.59, 0.59}  & \cellcolor[rgb]{0.59, 0.59, 0.59} \\ 
\selfcircle{6} & \cellcolor[rgb]{0.59, 0.59, 0.59}  & \cellcolor[rgb]{0.59, 0.59, 0.59}  & \cellcolor[rgb]{0.59, 0.59, 0.59}  & \cellcolor[rgb]{0.59, 0.59, 0.59}  & \cellcolor[rgb]{0.59, 0.59, 0.59}  & \cellcolor[rgb]{1.00, 1.00, 1.00}  & \cellcolor[rgb]{0.59, 0.59, 0.59} \\ 
\selfcircle{7} & \cellcolor[rgb]{0.59, 0.59, 0.59}  & \cellcolor[rgb]{0.59, 0.59, 0.59}  & \cellcolor[rgb]{0.59, 0.59, 0.59}  & \cellcolor[rgb]{0.59, 0.59, 0.59}  & \cellcolor[rgb]{0.59, 0.59, 0.59}  & \cellcolor[rgb]{0.59, 0.59, 0.59}  & \cellcolor[rgb]{1.00, 1.00, 1.00} \\ 
\bottomrule
\end{tabular}
}
\label{tab:app-sig-4}
\end{table}

From Tables~\ref{tab:app-sig-3} to ~\ref{tab:app-sig-4}, we observe that the best configuration \selfcircle{7} was significantly different from other configurations on most of test sets. The two exceptional cases are \selfcircle{7} versus \selfcircle{6} on test sets {\setELAHID} and {\setEDFHID}. Note that, \selfcircle{7}  was significantly worse than others on {\setELAII}. As we hypothesized in the main paper, other data augmentation techniques such as codec may be helpful to improve the performance on the ASVspoof2021 LA data.

\clearpage
\newpage

\subsection{Additional Results on Score Distributions}

\begin{figure}[h!]
\centering
\includegraphics[trim=0 20 0 10, width=\columnwidth]{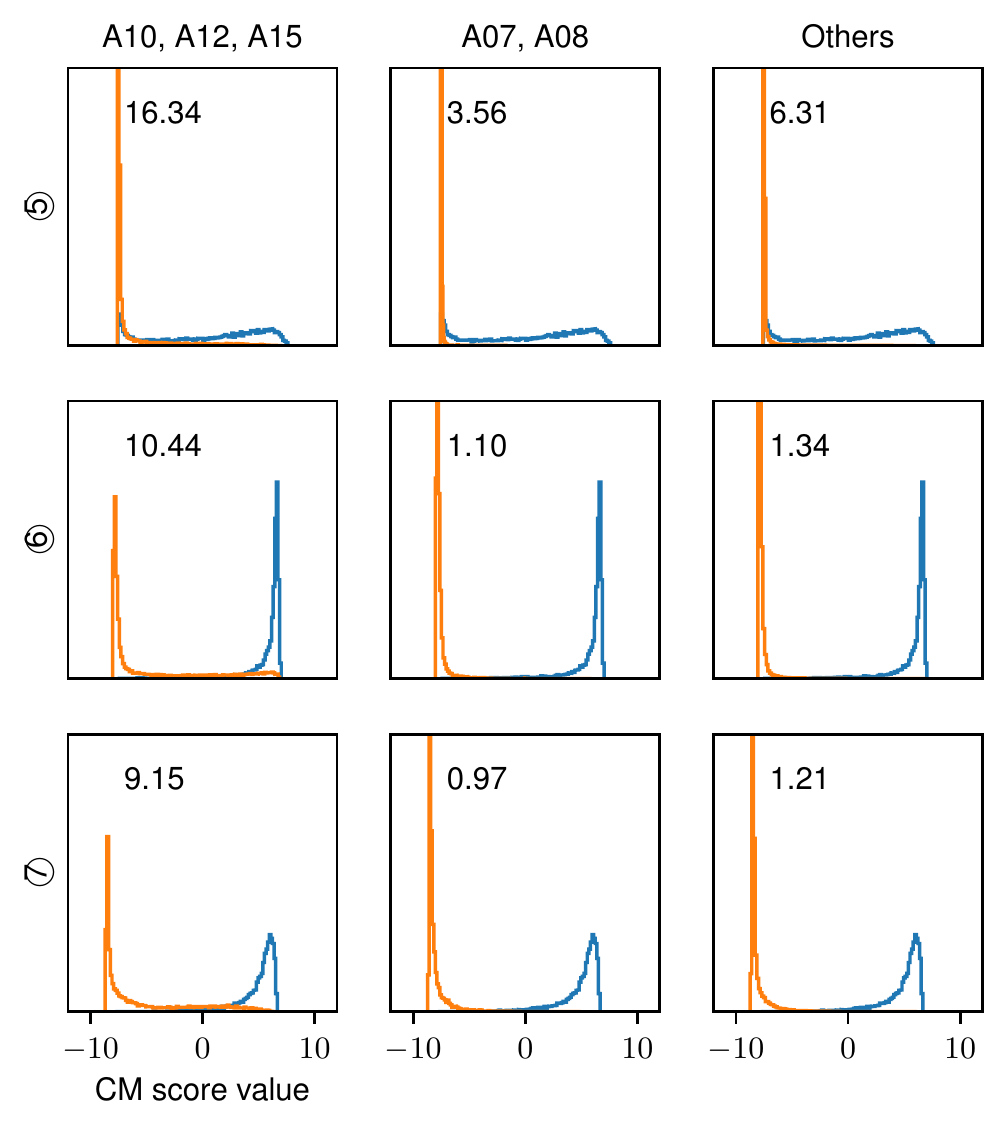}
\vspace{-2mm}
\caption{Score distributions of \selfcircle{5}, \selfcircle{6}, and \selfcircle{7} on \textcolor{blue}{bona fide} and \textcolor{orange}{spoofed} trials in {\setELATRIM}. Number in each sub-figure is EER (\%).} 
\label{fig:appscoredist1}
\end{figure}

\begin{figure}[t!]
\centering
\includegraphics[trim=0 20 0 10, width=\columnwidth]{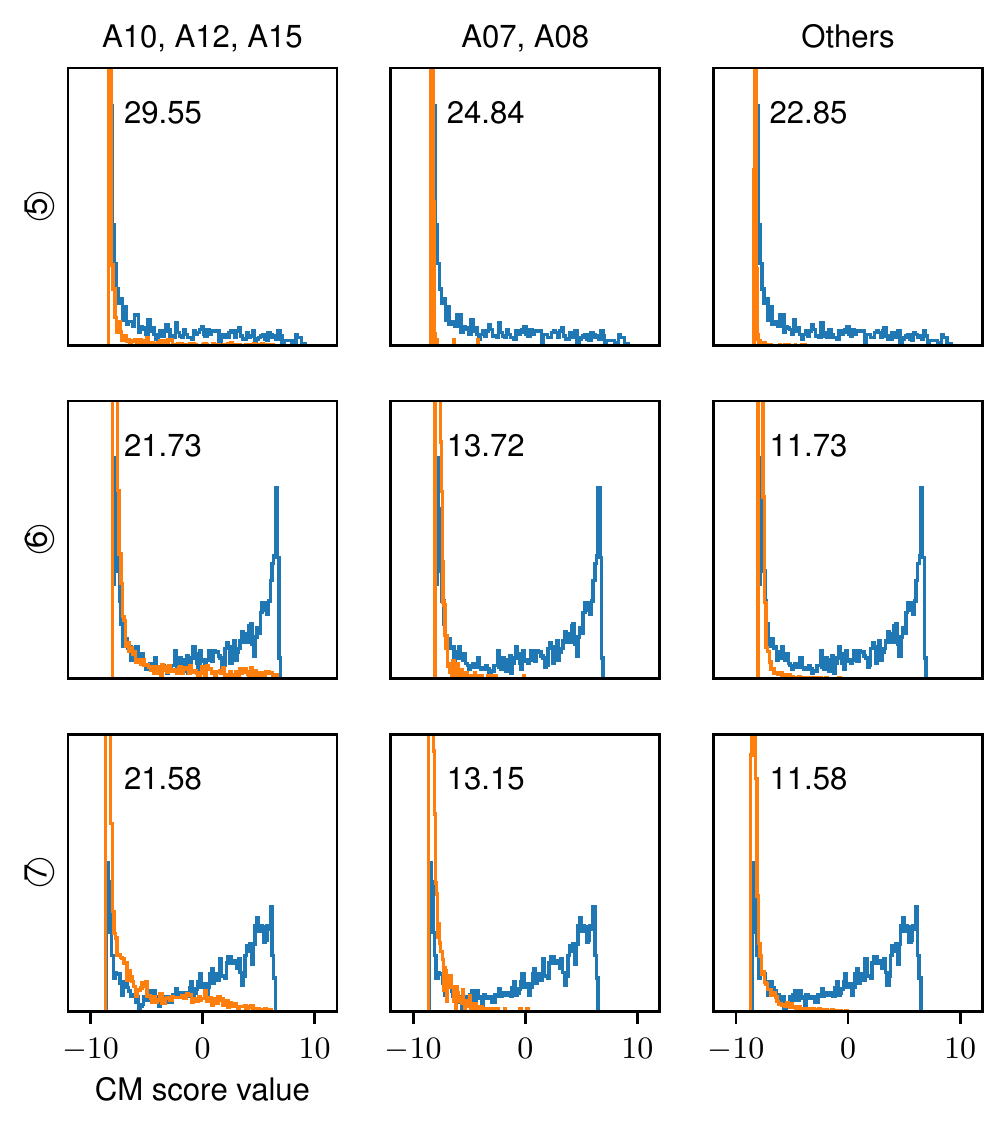}
\vspace{-2mm}
\caption{Score distributions of \selfcircle{5}, \selfcircle{6}, and \selfcircle{7} on \textcolor{blue}{bona fide} and \textcolor{orange}{spoofed} trials in {\setELAHID}. Number in each sub-figure is EER (\%).} 
\label{fig:appscoredist2}
\end{figure}

\begin{figure}[t!]
\centering
\includegraphics[trim=0 20 0 10, width=\columnwidth]{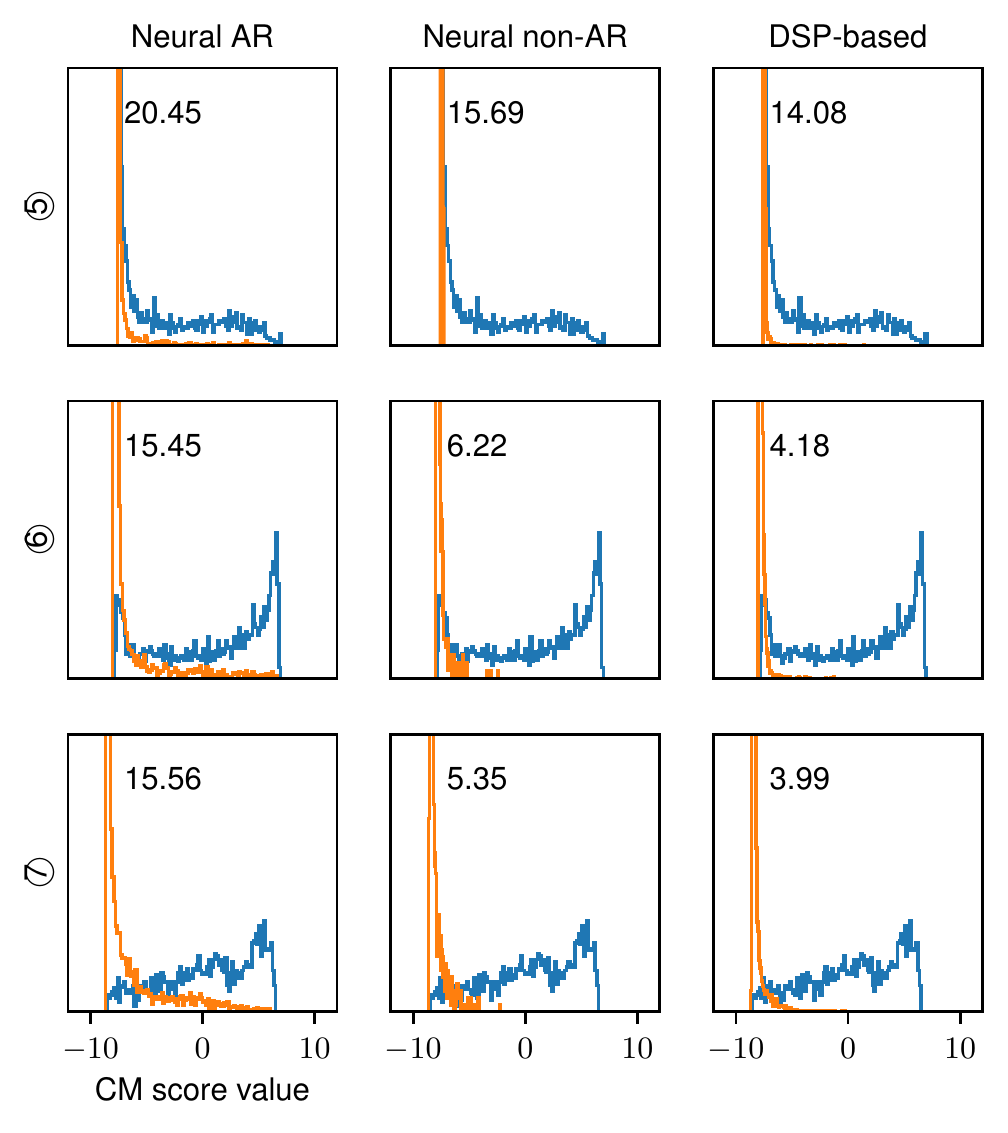}
\vspace{-2mm}
\caption{Score distributions of \selfcircle{5}, \selfcircle{6}, and \selfcircle{7} on \textcolor{blue}{bona fide} and \textcolor{orange}{spoofed} trials in {\setEDFHID}. Number in each sub-figure is EER (\%).} 
\label{fig:appscoredist3}
\end{figure}

Figures \ref{fig:appscoredist1} to \ref{fig:appscoredist3} show the score distributions of the three CMs \selfcircle{5}, \selfcircle{6}, and \selfcircle{7} on {\setELATRIM}, {\setELAHID}, and {\setEDFHID}. For {\setELATRIM} and {\setELAHID}, we group spoofing attack A10, A12, and A15 because they used neural AR vocoders. A07 and A08 used neural non-AR vocoders.

As we described in Section \label{sec:exp3}, the CMs performed worse on spoofed trials with neural AR vocoders since the CM training data does not cover similar vocoders. While other DSP-based vocoders are also unseen, the CM seemed to be generalizable.

\clearpage
\newpage

\subsection{Additional Results on Comparing Vocoded Training Sets}
\subsubsection{Additional CMs and Training Recipes}

In Section~\ref{sec:exp} we reported the results of different vocoded training sets on the SSL-based CM. During the experiments, we in fact compared three different types of CMs:
\begin{itemize}
\item {\mA}: a two-staged CM with front- and back-ends. The front end extracts linear frequency cepstrum coefficients (LFCC) from the input audio, and the back end is a light-convolution-neural-network (LCNN)-based classifier. Following the design in \citeapp{wang2021comparative},  {\mA}  added bi-directional recurrent layers and global average pooling after the LCNN to boost the performance. 
\item {\mB}: an end-to-end DNN-based CM that produces a waveform as input and produces a detection score. It contains graph attention networks and advanced pooling layers. {\mB} used the official implementation \citeapp{jung2022aasist}.
\item {\mC}: the SSL-based CM reported in Section~\ref{sec:exp} of the paper.
\end{itemize}
The three CMs were selected because they cover different categories of CMs in many existing works \citeapp{Wang2022}. {\mA} represents the common CMs with a DSP-based feature extractor and a DNN-based classifier. {\mB} directly processes waveform without the hardwired feature extractor, which represents the so-called end-to-end CM. It has achieved the lowest EER on the ASVspoof 2019 LA database \citeapp{jung2022aasist}. {\mC} also avoids conventional DSP features, but it leverages a pre-trained SSL speech model for feature extraction, a method showing competitive results on multiple databases \citeapp{wang2021investigating, tak2022automatic}.

The training recipe for {\mC} has been explained in Section~\ref{sec:exp} of the paper. The recipe for {\mA} and {\mB} are similar to that of {\mC} except that the learning rate was initialized to $3\times10^{-4}$ and multiplied with $0.5$ every ten epochs. The training process was stopped when the loss on the development set did not improve within 30 epochs. The batch size was 64.  All the training data were truncated into segments with a maximum duration of 4 seconds. During inference, the whole trial was processed without truncation. Similar to {\mC}, the other two CMs were also trained and evaluated for three rounds, and we report the averaged EERs.

In addition to the test sets reported in the main paper, we also test the results on the ASVspoof 2015 evaluation set {\setELAold} \citeapp{wu2015asvspoof}. We also report the EER on {\setTLAshort} for a sanity check.

\subsubsection{Results}

The results EERs are listed in Table~\ref{tab:app-exp-1}.  
First, when training on {\setTLAshort},  {\mA} and {\mB}'s EERs on the {\setELA} are close to the numbers reported in the literature \citeapp{wang2021investigating, jung2022aasist}. Thus, the implementation was properly set following the original literature.  However, both {\mA} and {\mB} obtained much higher EERs on other test sets, suggesting poor generalization across different test data. This result is also consistent with that in \citeapp{wang2021investigating}.  

\begin{table}[!t]
\caption{Addition results of test set EERs (\%) for Experiment 1. Each column corresponds to results of CM trained on one training set. Darker cell color indicates higher EER value. Some results in column {\setTWFshort} are removed due to data overlapping (see Section~\ref{sec:database}). Results of {\mC} are copied from Table~\ref{tab:exp-1} but have additional results on {\setTLAshort} and {\setELAold}.}
\begin{center}
\footnotesize
\vspace{-5mm}
\setlength{\tabcolsep}{4pt}
\resizebox{\columnwidth}{!}
{
\begin{tabular}{clcccccc}
\toprule
  &   &  \setTLA  &  \setTWF  &  \setTVI  & \setTVIII & \setTVIV  &  \setTVV \\ 
\midrule
\multirow{11}{*}{\rotatebox{90}{\mA}}          & \setTLAshort & \cellcolor[rgb]{1.00, 1.00, 1.00} 0.10 & \cellcolor[rgb]{0.69, 0.69, 0.69} 41.69 & \cellcolor[rgb]{0.93, 0.93, 0.93} 14.25 & \cellcolor[rgb]{0.70, 0.70, 0.70} 40.72 & \cellcolor[rgb]{0.87, 0.87, 0.87} 22.83 & \cellcolor[rgb]{0.83, 0.83, 0.83} 28.08\\ 
  &  \setELAold  & \cellcolor[rgb]{0.79, 0.79, 0.79} 32.42 & \cellcolor[rgb]{0.86, 0.86, 0.86} 23.82 & \cellcolor[rgb]{0.59, 0.59, 0.59} 49.89 & \cellcolor[rgb]{0.82, 0.82, 0.82} 28.34 & \cellcolor[rgb]{0.86, 0.86, 0.86} 23.90 & \cellcolor[rgb]{0.74, 0.74, 0.74} 37.60\\ 
  \cmidrule{2-8}
  &   \setELA   & \cellcolor[rgb]{0.99, 0.99, 0.99} 3.32 & \cellcolor[rgb]{0.59, 0.59, 0.59} 49.80 & \cellcolor[rgb]{0.86, 0.86, 0.86} 23.52 & \cellcolor[rgb]{0.70, 0.70, 0.70} 40.79 & \cellcolor[rgb]{0.74, 0.74, 0.74} 37.71 & \cellcolor[rgb]{0.81, 0.81, 0.81} 29.68\\ 
  &  \setELAII  & \cellcolor[rgb]{0.86, 0.86, 0.86} 23.38 & \cellcolor[rgb]{0.59, 0.59, 0.59} 57.18 & \cellcolor[rgb]{0.75, 0.75, 0.75} 36.43 & \cellcolor[rgb]{0.59, 0.59, 0.59} 62.69 & \cellcolor[rgb]{0.60, 0.60, 0.60} 49.17 & \cellcolor[rgb]{0.59, 0.59, 0.59} 49.84\\ 
  &   \setEDF   & \cellcolor[rgb]{0.81, 0.81, 0.81} 29.45 & \cellcolor[rgb]{0.62, 0.62, 0.62} 47.40 & \cellcolor[rgb]{0.71, 0.71, 0.71} 40.27 & \cellcolor[rgb]{0.59, 0.59, 0.59} 52.76 & \cellcolor[rgb]{0.62, 0.62, 0.62} 47.22 & \cellcolor[rgb]{0.59, 0.59, 0.59} 51.76\\ 
%  &   \setGA    & \cellcolor[rgb]{0.86, 0.86, 0.86} 24.41 & \cellcolor[rgb]{0.59, 0.59, 0.59} 52.49 & \cellcolor[rgb]{0.75, 0.75, 0.75} 37.03 & \cellcolor[rgb]{0.59, 0.59, 0.59} 55.17 & \cellcolor[rgb]{0.60, 0.60, 0.60} 49.61 & \cellcolor[rgb]{0.64, 0.64, 0.64} 45.61\\ 
  \cmidrule{2-8}
  & \setELATRIM & \cellcolor[rgb]{0.88, 0.88, 0.88} 21.18 & \cellcolor[rgb]{0.71, 0.71, 0.71} 40.10 & \cellcolor[rgb]{0.78, 0.78, 0.78} 33.23 & \cellcolor[rgb]{0.69, 0.69, 0.69} 41.97 & \cellcolor[rgb]{0.70, 0.70, 0.70} 40.94 & \cellcolor[rgb]{0.71, 0.71, 0.71} 40.56\\ 
  & \setELAHID  & \cellcolor[rgb]{0.72, 0.72, 0.72} 39.27 & \cellcolor[rgb]{0.60, 0.60, 0.60} 49.53 & \cellcolor[rgb]{0.65, 0.65, 0.65} 44.59 & \cellcolor[rgb]{0.60, 0.60, 0.60} 49.07 & \cellcolor[rgb]{0.66, 0.66, 0.66} 44.18 & \cellcolor[rgb]{0.60, 0.60, 0.60} 49.52\\ 
  & \setEDFHID  & \cellcolor[rgb]{0.76, 0.76, 0.76} 35.86 & \cellcolor[rgb]{0.65, 0.65, 0.65} 45.12 & \cellcolor[rgb]{0.67, 0.67, 0.67} 43.05 & \cellcolor[rgb]{0.61, 0.61, 0.61} 48.52 & \cellcolor[rgb]{0.65, 0.65, 0.65} 45.18 & \cellcolor[rgb]{0.59, 0.59, 0.59} 49.81\\ 
  &  \setEWFE   & \cellcolor[rgb]{0.64, 0.64, 0.64} 45.85 & - & \cellcolor[rgb]{0.87, 0.87, 0.87} 22.17 & \cellcolor[rgb]{0.88, 0.88, 0.88} 21.15 & \cellcolor[rgb]{0.93, 0.93, 0.93} 14.38 & \cellcolor[rgb]{0.84, 0.84, 0.84} 26.16\\ 
  &  \setEDFE   & \cellcolor[rgb]{0.59, 0.59, 0.59} 72.19 & \cellcolor[rgb]{0.59, 0.59, 0.59} 91.28 & \cellcolor[rgb]{0.59, 0.59, 0.59} 84.33 & \cellcolor[rgb]{0.64, 0.64, 0.64} 45.93 & \cellcolor[rgb]{0.59, 0.59, 0.59} 61.38 & \cellcolor[rgb]{0.82, 0.82, 0.82} 28.60\\ 
%  &   \setGB    & \cellcolor[rgb]{0.60, 0.60, 0.60} 49.19 & - & \cellcolor[rgb]{0.61, 0.61, 0.61} 47.86 & \cellcolor[rgb]{0.69, 0.69, 0.69} 41.63 & \cellcolor[rgb]{0.75, 0.75, 0.75} 37.08 & \cellcolor[rgb]{0.74, 0.74, 0.74} 37.28\\ 
  \cmidrule{2-8}
  &  \setGAll   & \cellcolor[rgb]{0.74, 0.74, 0.74} 37.68 & - & \cellcolor[rgb]{0.66, 0.66, 0.66} 43.90 & \cellcolor[rgb]{0.59, 0.59, 0.59} 51.16 & \cellcolor[rgb]{0.63, 0.63, 0.63} 46.26 & \cellcolor[rgb]{0.59, 0.59, 0.59} 51.35\\ 
\midrule
\multirow{11}{*}{\rotatebox{90}{\mB}}         & \setTLAshort & \cellcolor[rgb]{1.00, 1.00, 1.00} 0.66 & \cellcolor[rgb]{0.69, 0.69, 0.69} 41.50 & \cellcolor[rgb]{0.62, 0.62, 0.62} 47.58 & \cellcolor[rgb]{0.63, 0.63, 0.63} 46.46 & \cellcolor[rgb]{0.76, 0.76, 0.76} 34.97 & \cellcolor[rgb]{0.68, 0.68, 0.68} 42.46\\ 
  &  \setELAold  & \cellcolor[rgb]{0.99, 0.99, 0.99} 3.23 & \cellcolor[rgb]{0.82, 0.82, 0.82} 28.33 & \cellcolor[rgb]{0.62, 0.62, 0.62} 46.94 & \cellcolor[rgb]{0.82, 0.82, 0.82} 28.75 & \cellcolor[rgb]{0.72, 0.72, 0.72} 38.94 & \cellcolor[rgb]{0.68, 0.68, 0.68} 42.51\\ 
  \cmidrule{2-8}
    &   \setELA   & \cellcolor[rgb]{0.99, 0.99, 0.99} 1.54 & \cellcolor[rgb]{0.59, 0.59, 0.59} 53.64 & \cellcolor[rgb]{0.59, 0.59, 0.59} 61.13 & \cellcolor[rgb]{0.59, 0.59, 0.59} 51.57 & \cellcolor[rgb]{0.76, 0.76, 0.76} 35.05 & \cellcolor[rgb]{0.72, 0.72, 0.72} 39.79\\ 
  &  \setELAII  & \cellcolor[rgb]{0.95, 0.95, 0.95} 10.38 & \cellcolor[rgb]{0.59, 0.59, 0.59} 52.30 & \cellcolor[rgb]{0.59, 0.59, 0.59} 59.69 & \cellcolor[rgb]{0.59, 0.59, 0.59} 50.71 & \cellcolor[rgb]{0.75, 0.75, 0.75} 37.11 & \cellcolor[rgb]{0.68, 0.68, 0.68} 42.30\\ 
  &   \setEDF   & \cellcolor[rgb]{0.91, 0.91, 0.91} 16.99 & \cellcolor[rgb]{0.67, 0.67, 0.67} 43.61 & \cellcolor[rgb]{0.59, 0.59, 0.59} 54.52 & \cellcolor[rgb]{0.64, 0.64, 0.64} 45.53 & \cellcolor[rgb]{0.75, 0.75, 0.75} 36.18 & \cellcolor[rgb]{0.71, 0.71, 0.71} 40.53\\ 
%  &   \setGA    & \cellcolor[rgb]{0.94, 0.94, 0.94} 13.20 & \cellcolor[rgb]{0.59, 0.59, 0.59} 50.45 & \cellcolor[rgb]{0.59, 0.59, 0.59} 57.33 & \cellcolor[rgb]{0.60, 0.60, 0.60} 49.03 & \cellcolor[rgb]{0.75, 0.75, 0.75} 36.73 & \cellcolor[rgb]{0.70, 0.70, 0.70} 41.21\\ 
    \cmidrule{2-8}
  & \setELATRIM & \cellcolor[rgb]{0.90, 0.90, 0.90} 17.74 & \cellcolor[rgb]{0.59, 0.59, 0.59} 53.71 & \cellcolor[rgb]{0.64, 0.64, 0.64} 45.91 & \cellcolor[rgb]{0.59, 0.59, 0.59} 52.16 & \cellcolor[rgb]{0.67, 0.67, 0.67} 43.45 & \cellcolor[rgb]{0.60, 0.60, 0.60} 48.99\\ 
  & \setELAHID  & \cellcolor[rgb]{0.85, 0.85, 0.85} 25.70 & \cellcolor[rgb]{0.59, 0.59, 0.59} 51.29 & \cellcolor[rgb]{0.61, 0.61, 0.61} 48.39 & \cellcolor[rgb]{0.59, 0.59, 0.59} 50.67 & \cellcolor[rgb]{0.66, 0.66, 0.66} 44.21 & \cellcolor[rgb]{0.61, 0.61, 0.61} 47.96\\ 
  & \setEDFHID  & \cellcolor[rgb]{0.88, 0.88, 0.88} 20.76 & \cellcolor[rgb]{0.60, 0.60, 0.60} 49.38 & \cellcolor[rgb]{0.62, 0.62, 0.62} 46.97 & \cellcolor[rgb]{0.60, 0.60, 0.60} 49.41 & \cellcolor[rgb]{0.69, 0.69, 0.69} 41.86 & \cellcolor[rgb]{0.63, 0.63, 0.63} 46.85\\ 
  &  \setEWFE   & \cellcolor[rgb]{0.66, 0.66, 0.66} 43.88 & - & \cellcolor[rgb]{0.68, 0.68, 0.68} 42.62 & \cellcolor[rgb]{0.64, 0.64, 0.64} 46.01 & \cellcolor[rgb]{0.69, 0.69, 0.69} 41.80 & \cellcolor[rgb]{0.66, 0.66, 0.66} 44.21\\ 
  &  \setEDFE   & \cellcolor[rgb]{0.62, 0.62, 0.62} 46.94 & \cellcolor[rgb]{0.62, 0.62, 0.62} 46.97 & \cellcolor[rgb]{0.59, 0.59, 0.59} 50.97 & \cellcolor[rgb]{0.59, 0.59, 0.59} 56.05 & \cellcolor[rgb]{0.59, 0.59, 0.59} 55.70 & \cellcolor[rgb]{0.59, 0.59, 0.59} 58.68\\ 
%  &   \setGB    & \cellcolor[rgb]{0.76, 0.76, 0.76} 35.75 & - & \cellcolor[rgb]{0.59, 0.59, 0.59} 55.74 & \cellcolor[rgb]{0.59, 0.59, 0.59} 57.60 & \cellcolor[rgb]{0.59, 0.59, 0.59} 50.18 & \cellcolor[rgb]{0.59, 0.59, 0.59} 52.82\\ 
    \cmidrule{2-8}
  &  \setGAll   & \cellcolor[rgb]{0.85, 0.85, 0.85} 25.10 & - & \cellcolor[rgb]{0.59, 0.59, 0.59} 59.07 & \cellcolor[rgb]{0.59, 0.59, 0.59} 57.87 & \cellcolor[rgb]{0.68, 0.68, 0.68} 42.69 & \cellcolor[rgb]{0.61, 0.61, 0.61} 48.41\\ 
\midrule
\multirow{11}{*}{\rotatebox{90}{\mC}}      & \setTLAshort & \cellcolor[rgb]{1.00, 1.00, 1.00} 0.06 & \cellcolor[rgb]{0.86, 0.86, 0.86} 23.53 & \cellcolor[rgb]{0.96, 0.96, 0.96} 7.57 & \cellcolor[rgb]{0.97, 0.97, 0.97} 6.62 & \cellcolor[rgb]{0.98, 0.98, 0.98} 4.10 & \cellcolor[rgb]{0.99, 0.99, 0.99} 1.47\\ 
  &  \setELAold  & \cellcolor[rgb]{1.00, 1.00, 1.00} 0.18 & \cellcolor[rgb]{0.94, 0.94, 0.94} 12.34 & \cellcolor[rgb]{0.59, 0.59, 0.59} 58.06 & \cellcolor[rgb]{0.97, 0.97, 0.97} 5.64 & \cellcolor[rgb]{1.00, 1.00, 1.00} 1.13 & \cellcolor[rgb]{1.00, 1.00, 1.00} 1.09\\ 
  \cmidrule{2-8}
    &   \setELA   & \cellcolor[rgb]{0.99, 0.99, 0.99} 2.98 & \cellcolor[rgb]{0.66, 0.66, 0.66} 44.48 & \cellcolor[rgb]{0.97, 0.97, 0.97} 5.78 & \cellcolor[rgb]{0.98, 0.98, 0.98} 5.32 & \cellcolor[rgb]{0.96, 0.96, 0.96} 8.74 & \cellcolor[rgb]{0.98, 0.98, 0.98} 4.36\\ 
  &  \setELAII  & \cellcolor[rgb]{0.96, 0.96, 0.96} 7.53 & \cellcolor[rgb]{0.69, 0.69, 0.69} 41.57 & \cellcolor[rgb]{0.84, 0.84, 0.84} 26.30 & \cellcolor[rgb]{0.90, 0.90, 0.90} 17.98 & \cellcolor[rgb]{0.89, 0.89, 0.89} 19.29 & \cellcolor[rgb]{0.86, 0.86, 0.86} 24.39\\ 
  &   \setEDF   & \cellcolor[rgb]{0.97, 0.97, 0.97} 6.67 & \cellcolor[rgb]{0.86, 0.86, 0.86} 24.26 & \cellcolor[rgb]{0.94, 0.94, 0.94} 11.95 & \cellcolor[rgb]{0.95, 0.95, 0.95} 11.54 & \cellcolor[rgb]{0.96, 0.96, 0.96} 9.71 & \cellcolor[rgb]{0.94, 0.94, 0.94} 13.31\\ 
%  &   \setGA    & \cellcolor[rgb]{0.97, 0.97, 0.97} 6.20 & \cellcolor[rgb]{0.80, 0.80, 0.80} 30.57 & \cellcolor[rgb]{0.92, 0.92, 0.92} 15.93 & \cellcolor[rgb]{0.94, 0.94, 0.94} 12.73 & \cellcolor[rgb]{0.94, 0.94, 0.94} 12.31 & \cellcolor[rgb]{0.92, 0.92, 0.92} 15.54\\ 
\cmidrule{2-8}
  & \setELATRIM & \cellcolor[rgb]{0.92, 0.92, 0.92} 15.56 & \cellcolor[rgb]{0.80, 0.80, 0.80} 31.62 & \cellcolor[rgb]{0.86, 0.86, 0.86} 23.29 & \cellcolor[rgb]{0.92, 0.92, 0.92} 16.16 & \cellcolor[rgb]{0.92, 0.92, 0.92} 14.99 & \cellcolor[rgb]{0.96, 0.96, 0.96} 9.52\\ 
  & \setELAHID  & \cellcolor[rgb]{0.82, 0.82, 0.82} 28.80 & \cellcolor[rgb]{0.83, 0.83, 0.83} 27.60 & \cellcolor[rgb]{0.82, 0.82, 0.82} 28.30 & \cellcolor[rgb]{0.89, 0.89, 0.89} 19.49 & \cellcolor[rgb]{0.90, 0.90, 0.90} 17.62 & \cellcolor[rgb]{0.88, 0.88, 0.88} 21.43\\ 
  & \setEDFHID  & \cellcolor[rgb]{0.86, 0.86, 0.86} 23.62 & \cellcolor[rgb]{0.84, 0.84, 0.84} 26.18 & \cellcolor[rgb]{0.87, 0.87, 0.87} 22.01 & \cellcolor[rgb]{0.93, 0.93, 0.93} 13.92 & \cellcolor[rgb]{0.94, 0.94, 0.94} 13.50 & \cellcolor[rgb]{0.91, 0.91, 0.91} 16.99\\ 
  &  \setEWFE   & \cellcolor[rgb]{0.92, 0.92, 0.92} 15.76 & - & \cellcolor[rgb]{0.72, 0.72, 0.72} 39.27 & \cellcolor[rgb]{0.77, 0.77, 0.77} 34.05 & \cellcolor[rgb]{0.91, 0.91, 0.91} 17.10 & \cellcolor[rgb]{0.95, 0.95, 0.95} 10.89\\ 
  &  \setEDFE   & \cellcolor[rgb]{0.84, 0.84, 0.84} 26.65 & \cellcolor[rgb]{0.89, 0.89, 0.89} 19.98 & \cellcolor[rgb]{0.70, 0.70, 0.70} 41.06 & \cellcolor[rgb]{0.75, 0.75, 0.75} 36.46 & \cellcolor[rgb]{0.87, 0.87, 0.87} 22.26 & \cellcolor[rgb]{0.89, 0.89, 0.89} 19.45\\ 
%   &   \setGB    & \cellcolor[rgb]{0.83, 0.83, 0.83} 27.84 & - & \cellcolor[rgb]{0.66, 0.66, 0.66} 44.47 & \cellcolor[rgb]{0.65, 0.65, 0.65} 45.07 & \cellcolor[rgb]{0.80, 0.80, 0.80} 30.90 & \cellcolor[rgb]{0.86, 0.86, 0.86} 23.23\\ 
\cmidrule{2-8}
  &  \setGAll   & \cellcolor[rgb]{0.93, 0.93, 0.93} 14.24 & - & \cellcolor[rgb]{0.75, 0.75, 0.75} 36.57 & \cellcolor[rgb]{0.71, 0.71, 0.71} 39.95 & \cellcolor[rgb]{0.89, 0.89, 0.89} 19.39 & \cellcolor[rgb]{0.92, 0.92, 0.92} 16.35\\ 
\bottomrule
\end{tabular}
}
\end{center}
\label{tab:app-exp-1}
\end{table}

\begin{table*}[!t]
\caption{Additional results on contrastive feature loss (EERs \%). Each EER value is averaged over results from three independent training-evaluation rounds. Darker cell color indicates higher EER value.  
Systems IDs \selfcircle{1} - \selfcircle{7} have been reported in Table~\ref{tab:exp-2} in main paper. Other entries are additional results.
}
\vspace{-4mm}
\begin{center}
\resizebox{\textwidth}{!}{
\begin{tabular}{ccccccccccccccc}
\toprule
  &   \multicolumn{1}{l}{Training criterion}  & \multicolumn{4}{c}{$\mathcal{L}_{\text{CE}}$ }   & \multicolumn{9}{c}{$\mathcal{L}_{\text{CE}} + \mathcal{L}_{\text{CF}}$ (one additional view)}   \\
  \cmidrule(lr){3-6}\cmidrule(lr){7-15} 
   &  \multicolumn{1}{l}{Data augmentation} &  \multicolumn{2}{c}{$\times$} & \multicolumn{2}{c}{RawBoost} & \multicolumn{3}{c}{RawBoost} &  \multicolumn{3}{c}{Codec} &  \multicolumn{3}{c}{Freq.Mask}  \\
  \cmidrule(lr){3-4}\cmidrule(lr){5-6}\cmidrule(lr){7-9} \cmidrule(lr){10-12} \cmidrule(lr){13-15}
  &   \multicolumn{1}{l}{Training set}    & \setTLA & \setTVV & \setTLA & \setTVV & \setTLA & \setTVV & \setTVV & \setTLA & \setTVV & \setTVV & \setTLA & \setTVV & \setTVV \\ 
  & \multicolumn{1}{l}{Bona-spoof paired}   &  $\times$  &  $\times$  &  $\times$  &  $\times$  &  $\times$  &  $\times$ &  $\checkmark$ &  $\times$  &  $\times$ &  $\checkmark$ &  $\times$  &  $\times$ &  $\checkmark$\\
 & \multicolumn{1}{l}{ID in Table~\ref{tab:exp-2}}  & \selfcircle{1} & \selfcircle{2} & \selfcircle{3} & \selfcircle{4} & \selfcircle{5} & \selfcircle{6} & \selfcircle{7} \\
\midrule
\multirow{9}{*}{\rotatebox{90}{\mC}}       & \setELAold  & \cellcolor[rgb]{1.00, 1.00, 1.00} 0.18 & \cellcolor[rgb]{1.00, 1.00, 1.00} 1.09 & \cellcolor[rgb]{1.00, 1.00, 1.00} 0.54 & \cellcolor[rgb]{1.00, 1.00, 1.00} 0.68 & \cellcolor[rgb]{1.00, 1.00, 1.00} 0.59 & \cellcolor[rgb]{1.00, 1.00, 1.00} 0.37 & \cellcolor[rgb]{1.00, 1.00, 1.00} 0.55 & \cellcolor[rgb]{1.00, 1.00, 1.00} 0.72 & \cellcolor[rgb]{1.00, 1.00, 1.00} 0.23 & \cellcolor[rgb]{1.00, 1.00, 1.00} 0.38 & \cellcolor[rgb]{1.00, 1.00, 1.00} 0.40 & \cellcolor[rgb]{1.00, 1.00, 1.00} 1.60 & \cellcolor[rgb]{1.00, 1.00, 1.00} 1.36\\ 
\cmidrule{2-15}
  &   \setELA   & \cellcolor[rgb]{0.99, 0.99, 0.99} 2.98 & \cellcolor[rgb]{0.98, 0.98, 0.98} 4.36 & \cellcolor[rgb]{1.00, 1.00, 1.00} 0.22 & \cellcolor[rgb]{0.99, 0.99, 0.99} 3.46 & \cellcolor[rgb]{1.00, 1.00, 1.00} 0.21 & \cellcolor[rgb]{1.00, 1.00, 1.00} 2.63 & \cellcolor[rgb]{1.00, 1.00, 1.00} 2.21 & \cellcolor[rgb]{1.00, 1.00, 1.00} 2.65 & \cellcolor[rgb]{0.99, 0.99, 0.99} 3.47 & \cellcolor[rgb]{0.99, 0.99, 0.99} 3.87 & \cellcolor[rgb]{1.00, 1.00, 1.00} 1.35 & \cellcolor[rgb]{0.98, 0.98, 0.98} 5.05 & \cellcolor[rgb]{1.00, 1.00, 1.00} 2.33\\ 
  &  \setELAII  & \cellcolor[rgb]{0.96, 0.96, 0.96} 7.53 & \cellcolor[rgb]{0.76, 0.76, 0.76} 24.39 & \cellcolor[rgb]{0.99, 0.99, 0.99} 3.63 & \cellcolor[rgb]{0.87, 0.87, 0.87} 16.55 & \cellcolor[rgb]{0.99, 0.99, 0.99} 3.30 & \cellcolor[rgb]{0.86, 0.86, 0.86} 16.67 & \cellcolor[rgb]{0.85, 0.85, 0.85} 17.90 & \cellcolor[rgb]{0.98, 0.98, 0.98} 4.29 & \cellcolor[rgb]{0.94, 0.94, 0.94} 9.54 & \cellcolor[rgb]{0.94, 0.94, 0.94} 10.56 & \cellcolor[rgb]{0.97, 0.97, 0.97} 6.13 & \cellcolor[rgb]{0.86, 0.86, 0.86} 17.34 & \cellcolor[rgb]{0.86, 0.86, 0.86} 17.48\\ 
  &   \setEDF   & \cellcolor[rgb]{0.97, 0.97, 0.97} 6.67 & \cellcolor[rgb]{0.90, 0.90, 0.90} 13.31 & \cellcolor[rgb]{0.99, 0.99, 0.99} 3.65 & \cellcolor[rgb]{0.94, 0.94, 0.94} 9.60 & \cellcolor[rgb]{0.99, 0.99, 0.99} 4.12 & \cellcolor[rgb]{0.96, 0.96, 0.96} 6.92 & \cellcolor[rgb]{0.98, 0.98, 0.98} 5.04 & \cellcolor[rgb]{0.97, 0.97, 0.97} 6.24 & \cellcolor[rgb]{0.97, 0.97, 0.97} 6.68 & \cellcolor[rgb]{0.97, 0.97, 0.97} 5.51 & \cellcolor[rgb]{0.97, 0.97, 0.97} 6.30 & \cellcolor[rgb]{0.90, 0.90, 0.90} 13.64 & \cellcolor[rgb]{0.92, 0.92, 0.92} 12.02\\ 
  \cmidrule{2-15}
%  &   \setGA    & \cellcolor[rgb]{0.97, 0.97, 0.97} 6.20 & \cellcolor[rgb]{0.88, 0.88, 0.88} 15.54 & \cellcolor[rgb]{0.99, 0.99, 0.99} 3.30 & \cellcolor[rgb]{0.93, 0.93, 0.93} 10.68 & \cellcolor[rgb]{0.99, 0.99, 0.99} 3.50 & \cellcolor[rgb]{0.94, 0.94, 0.94} 9.63 & \cellcolor[rgb]{0.95, 0.95, 0.95} 9.19 & \cellcolor[rgb]{0.98, 0.98, 0.98} 4.77 & \cellcolor[rgb]{0.96, 0.96, 0.96} 6.83 & \cellcolor[rgb]{0.97, 0.97, 0.97} 6.44 & \cellcolor[rgb]{0.97, 0.97, 0.97} 5.76 & \cellcolor[rgb]{0.91, 0.91, 0.91} 12.76 & \cellcolor[rgb]{0.91, 0.91, 0.91} 12.73\\ 
  & \setELATRIM & \cellcolor[rgb]{0.88, 0.88, 0.88} 15.56 & \cellcolor[rgb]{0.94, 0.94, 0.94} 9.52 & \cellcolor[rgb]{0.95, 0.95, 0.95} 9.16 & \cellcolor[rgb]{0.97, 0.97, 0.97} 6.09 & \cellcolor[rgb]{0.95, 0.95, 0.95} 9.00 & \cellcolor[rgb]{0.98, 0.98, 0.98} 4.48 & \cellcolor[rgb]{0.99, 0.99, 0.99} 3.79 & \cellcolor[rgb]{0.92, 0.92, 0.92} 11.45 & \cellcolor[rgb]{0.96, 0.96, 0.96} 7.51 & \cellcolor[rgb]{0.97, 0.97, 0.97} 5.98 & \cellcolor[rgb]{0.95, 0.95, 0.95} 8.72 & \cellcolor[rgb]{0.97, 0.97, 0.97} 5.73 & \cellcolor[rgb]{0.98, 0.98, 0.98} 5.26\\ 
  & \setELAHID  & \cellcolor[rgb]{0.68, 0.68, 0.68} 28.80 & \cellcolor[rgb]{0.80, 0.80, 0.80} 21.43 & \cellcolor[rgb]{0.81, 0.81, 0.81} 21.18 & \cellcolor[rgb]{0.83, 0.83, 0.83} 19.37 & \cellcolor[rgb]{0.72, 0.72, 0.72} 26.98 & \cellcolor[rgb]{0.88, 0.88, 0.88} 15.05 & \cellcolor[rgb]{0.89, 0.89, 0.89} 14.57 & \cellcolor[rgb]{0.80, 0.80, 0.80} 21.34 & \cellcolor[rgb]{0.91, 0.91, 0.91} 12.74 & \cellcolor[rgb]{0.92, 0.92, 0.92} 11.64 & \cellcolor[rgb]{0.72, 0.72, 0.72} 26.96 & \cellcolor[rgb]{0.86, 0.86, 0.86} 16.92 & \cellcolor[rgb]{0.87, 0.87, 0.87} 16.22\\ 
  & \setEDFHID  & \cellcolor[rgb]{0.77, 0.77, 0.77} 23.62 & \cellcolor[rgb]{0.86, 0.86, 0.86} 16.99 & \cellcolor[rgb]{0.90, 0.90, 0.90} 13.64 & \cellcolor[rgb]{0.89, 0.89, 0.89} 14.29 & \cellcolor[rgb]{0.86, 0.86, 0.86} 16.85 & \cellcolor[rgb]{0.96, 0.96, 0.96} 8.17 & \cellcolor[rgb]{0.96, 0.96, 0.96} 7.78 & \cellcolor[rgb]{0.89, 0.89, 0.89} 14.77 & \cellcolor[rgb]{0.94, 0.94, 0.94} 9.56 & \cellcolor[rgb]{0.95, 0.95, 0.95} 8.50 & \cellcolor[rgb]{0.84, 0.84, 0.84} 18.30 & \cellcolor[rgb]{0.89, 0.89, 0.89} 14.41 & \cellcolor[rgb]{0.89, 0.89, 0.89} 14.10\\ 
  &  \setEWFE   & \cellcolor[rgb]{0.88, 0.88, 0.88} 15.76 & \cellcolor[rgb]{0.93, 0.93, 0.93} 10.89 & \cellcolor[rgb]{0.73, 0.73, 0.73} 26.37 & \cellcolor[rgb]{0.96, 0.96, 0.96} 6.87 & \cellcolor[rgb]{0.76, 0.76, 0.76} 24.62 & \cellcolor[rgb]{0.99, 0.99, 0.99} 4.03 & \cellcolor[rgb]{1.00, 1.00, 1.00} 2.50 & \cellcolor[rgb]{0.75, 0.75, 0.75} 25.34 & \cellcolor[rgb]{0.96, 0.96, 0.96} 7.63 & \cellcolor[rgb]{0.98, 0.98, 0.98} 4.70 & \cellcolor[rgb]{0.74, 0.74, 0.74} 25.58 & \cellcolor[rgb]{0.90, 0.90, 0.90} 13.41 & \cellcolor[rgb]{0.95, 0.95, 0.95} 8.49\\ 
  &  \setEDFE   & \cellcolor[rgb]{0.72, 0.72, 0.72} 26.65 & \cellcolor[rgb]{0.83, 0.83, 0.83} 19.45 & \cellcolor[rgb]{0.87, 0.87, 0.87} 16.17 & \cellcolor[rgb]{0.92, 0.92, 0.92} 12.08 & \cellcolor[rgb]{0.86, 0.86, 0.86} 17.07 & \cellcolor[rgb]{0.95, 0.95, 0.95} 9.37 & \cellcolor[rgb]{0.96, 0.96, 0.96} 7.55 & \cellcolor[rgb]{0.83, 0.83, 0.83} 19.18 & \cellcolor[rgb]{0.88, 0.88, 0.88} 14.88 & \cellcolor[rgb]{0.91, 0.91, 0.91} 12.94 & \cellcolor[rgb]{0.86, 0.86, 0.86} 16.93 & \cellcolor[rgb]{0.88, 0.88, 0.88} 15.06 & \cellcolor[rgb]{0.92, 0.92, 0.92} 12.09\\ 
  \cmidrule{2-15}
%  &   \setGB    & \cellcolor[rgb]{0.70, 0.70, 0.70} 27.84 & \cellcolor[rgb]{0.77, 0.77, 0.77} 23.23 & \cellcolor[rgb]{0.59, 0.59, 0.59} 33.58 & \cellcolor[rgb]{0.82, 0.82, 0.82} 19.99 & \cellcolor[rgb]{0.60, 0.60, 0.60} 33.07 & \cellcolor[rgb]{0.84, 0.84, 0.84} 19.02 & \cellcolor[rgb]{0.89, 0.89, 0.89} 14.22 & \cellcolor[rgb]{0.71, 0.71, 0.71} 27.39 & \cellcolor[rgb]{0.87, 0.87, 0.87} 16.40 & \cellcolor[rgb]{0.88, 0.88, 0.88} 15.35 & \cellcolor[rgb]{0.69, 0.69, 0.69} 28.54 & \cellcolor[rgb]{0.85, 0.85, 0.85} 17.75 & \cellcolor[rgb]{0.85, 0.85, 0.85} 17.84\\ 
  &  \setGAll   & \cellcolor[rgb]{0.89, 0.89, 0.89} 14.24 & \cellcolor[rgb]{0.87, 0.87, 0.87} 16.35 & \cellcolor[rgb]{0.90, 0.90, 0.90} 13.12 & \cellcolor[rgb]{0.90, 0.90, 0.90} 13.13 & \cellcolor[rgb]{0.90, 0.90, 0.90} 13.68 & \cellcolor[rgb]{0.90, 0.90, 0.90} 13.15 & \cellcolor[rgb]{0.93, 0.93, 0.93} 11.27 & \cellcolor[rgb]{0.89, 0.89, 0.89} 14.79 & \cellcolor[rgb]{0.92, 0.92, 0.92} 11.95 & \cellcolor[rgb]{0.93, 0.93, 0.93} 10.85 & \cellcolor[rgb]{0.90, 0.90, 0.90} 13.91 & \cellcolor[rgb]{0.89, 0.89, 0.89} 14.57 & \cellcolor[rgb]{0.89, 0.89, 0.89} 14.42\\ 
\bottomrule
\\
\toprule
  &   \multicolumn{1}{l}{Training criterion}  & \multicolumn{9}{c}{$\mathcal{L}_{\text{CE}} + \mathcal{L}_{\text{CF}}$ (two or three additional views)}   \\
  \cmidrule(lr){3-11} 
   &  \multicolumn{1}{l}{Data augmentation} &  \multicolumn{3}{c}{RawBoost,Codec} &  \multicolumn{3}{c}{RawBoost,Freq.Mask} &  \multicolumn{3}{c}{RawBoost,Codec,Freq.M}  \\
  \cmidrule(lr){3-5}\cmidrule(lr){6-8} \cmidrule(lr){9-11}
  &   \multicolumn{1}{l}{Training set}   & \setTLA & \setTVV & \setTVV & \setTLA & \setTVV & \setTVV & \setTLA & \setTVV & \setTVV \\ 
  & \multicolumn{1}{l}{Bona-spoof paired}   & $\times$  &  $\times$ &  $\checkmark$ &  $\times$  &  $\times$ &  $\checkmark$ &  $\times$  &  $\times$ &  $\checkmark$\\
\midrule
 \multirow{9}{*}{\rotatebox{90}{\mC}}      & \setELAold  & \cellcolor[rgb]{1.00, 1.00, 1.00} 1.82 & \cellcolor[rgb]{1.00, 1.00, 1.00} 0.35 & \cellcolor[rgb]{1.00, 1.00, 1.00} 0.33 & \cellcolor[rgb]{1.00, 1.00, 1.00} 0.55 & \cellcolor[rgb]{1.00, 1.00, 1.00} 0.46 & \cellcolor[rgb]{1.00, 1.00, 1.00} 0.48 & \cellcolor[rgb]{1.00, 1.00, 1.00} 1.58 & \cellcolor[rgb]{1.00, 1.00, 1.00} 0.38 & \cellcolor[rgb]{1.00, 1.00, 1.00} 0.28\\ 
  \cmidrule{2-11}
    &   \setELA   & \cellcolor[rgb]{1.00, 1.00, 1.00} 0.90 & \cellcolor[rgb]{0.99, 0.99, 0.99} 3.63 & \cellcolor[rgb]{0.99, 0.99, 0.99} 3.33 & \cellcolor[rgb]{1.00, 1.00, 1.00} 0.15 & \cellcolor[rgb]{0.99, 0.99, 0.99} 3.24 & \cellcolor[rgb]{1.00, 1.00, 1.00} 2.33 & \cellcolor[rgb]{1.00, 1.00, 1.00} 0.61 & \cellcolor[rgb]{0.99, 0.99, 0.99} 4.02 & \cellcolor[rgb]{1.00, 1.00, 1.00} 2.28\\ 
  &  \setELAII  & \cellcolor[rgb]{1.00, 1.00, 1.00} 2.04 & \cellcolor[rgb]{0.95, 0.95, 0.95} 9.62 & \cellcolor[rgb]{0.94, 0.94, 0.94} 10.73 & \cellcolor[rgb]{0.99, 0.99, 0.99} 3.40 & \cellcolor[rgb]{0.87, 0.87, 0.87} 17.74 & \cellcolor[rgb]{0.84, 0.84, 0.84} 19.99 & \cellcolor[rgb]{1.00, 1.00, 1.00} 1.95 & \cellcolor[rgb]{0.94, 0.94, 0.94} 10.35 & \cellcolor[rgb]{0.95, 0.95, 0.95} 9.07\\ 
  &   \setEDF   & \cellcolor[rgb]{0.98, 0.98, 0.98} 5.66 & \cellcolor[rgb]{0.98, 0.98, 0.98} 5.52 & \cellcolor[rgb]{0.99, 0.99, 0.99} 4.01 & \cellcolor[rgb]{0.98, 0.98, 0.98} 4.58 & \cellcolor[rgb]{0.96, 0.96, 0.96} 8.20 & \cellcolor[rgb]{0.98, 0.98, 0.98} 5.61 & \cellcolor[rgb]{0.98, 0.98, 0.98} 4.93 & \cellcolor[rgb]{0.96, 0.96, 0.96} 7.15 & \cellcolor[rgb]{0.99, 0.99, 0.99} 3.34\\ 
%  &   \setGA    & \cellcolor[rgb]{0.99, 0.99, 0.99} 4.03 & \cellcolor[rgb]{0.97, 0.97, 0.97} 6.30 & \cellcolor[rgb]{0.98, 0.98, 0.98} 5.63 & \cellcolor[rgb]{0.99, 0.99, 0.99} 3.82 & \cellcolor[rgb]{0.94, 0.94, 0.94} 10.60 & \cellcolor[rgb]{0.94, 0.94, 0.94} 10.31 & \cellcolor[rgb]{0.99, 0.99, 0.99} 3.65 & \cellcolor[rgb]{0.96, 0.96, 0.96} 8.16 & \cellcolor[rgb]{0.98, 0.98, 0.98} 4.84\\ 
  \cmidrule{2-11}
  & \setELATRIM & \cellcolor[rgb]{0.95, 0.95, 0.95} 9.59 & \cellcolor[rgb]{0.98, 0.98, 0.98} 5.38 & \cellcolor[rgb]{0.98, 0.98, 0.98} 4.95 & \cellcolor[rgb]{0.96, 0.96, 0.96} 7.85 & \cellcolor[rgb]{0.98, 0.98, 0.98} 4.45 & \cellcolor[rgb]{0.99, 0.99, 0.99} 3.90 & \cellcolor[rgb]{0.96, 0.96, 0.96} 8.64 & \cellcolor[rgb]{0.98, 0.98, 0.98} 5.46 & \cellcolor[rgb]{0.98, 0.98, 0.98} 4.76\\ 
  & \setELAHID  & \cellcolor[rgb]{0.90, 0.90, 0.90} 14.10 & \cellcolor[rgb]{0.95, 0.95, 0.95} 9.78 & \cellcolor[rgb]{0.95, 0.95, 0.95} 9.66 & \cellcolor[rgb]{0.68, 0.68, 0.68} 31.23 & \cellcolor[rgb]{0.89, 0.89, 0.89} 15.29 & \cellcolor[rgb]{0.90, 0.90, 0.90} 15.03 & \cellcolor[rgb]{0.89, 0.89, 0.89} 15.85 & \cellcolor[rgb]{0.94, 0.94, 0.94} 10.60 & \cellcolor[rgb]{0.94, 0.94, 0.94} 10.54\\ 
  & \setEDFHID  & \cellcolor[rgb]{0.95, 0.95, 0.95} 9.83 & \cellcolor[rgb]{0.97, 0.97, 0.97} 6.13 & \cellcolor[rgb]{0.97, 0.97, 0.97} 6.11 & \cellcolor[rgb]{0.84, 0.84, 0.84} 20.25 & \cellcolor[rgb]{0.96, 0.96, 0.96} 8.40 & \cellcolor[rgb]{0.96, 0.96, 0.96} 8.43 & \cellcolor[rgb]{0.94, 0.94, 0.94} 10.39 & \cellcolor[rgb]{0.96, 0.96, 0.96} 7.33 & \cellcolor[rgb]{0.97, 0.97, 0.97} 6.50\\ 
  &  \setEWFE   & \cellcolor[rgb]{0.62, 0.62, 0.62} 34.49 & \cellcolor[rgb]{0.96, 0.96, 0.96} 7.88 & \cellcolor[rgb]{0.99, 0.99, 0.99} 2.86 & \cellcolor[rgb]{0.87, 0.87, 0.87} 17.00 & \cellcolor[rgb]{0.99, 0.99, 0.99} 4.33 & \cellcolor[rgb]{0.99, 0.99, 0.99} 3.39 & \cellcolor[rgb]{0.69, 0.69, 0.69} 30.44 & \cellcolor[rgb]{0.96, 0.96, 0.96} 8.68 & \cellcolor[rgb]{0.98, 0.98, 0.98} 4.63\\ 
  &  \setEDFE   & \cellcolor[rgb]{0.86, 0.86, 0.86} 18.47 & \cellcolor[rgb]{0.95, 0.95, 0.95} 8.75 & \cellcolor[rgb]{0.95, 0.95, 0.95} 9.22 & \cellcolor[rgb]{0.90, 0.90, 0.90} 14.84 & \cellcolor[rgb]{0.95, 0.95, 0.95} 8.88 & \cellcolor[rgb]{0.96, 0.96, 0.96} 8.66 & \cellcolor[rgb]{0.89, 0.89, 0.89} 15.30 & \cellcolor[rgb]{0.93, 0.93, 0.93} 11.60 & \cellcolor[rgb]{0.94, 0.94, 0.94} 10.21\\ 
%  &   \setGB    & \cellcolor[rgb]{0.59, 0.59, 0.59} 36.26 & \cellcolor[rgb]{0.88, 0.88, 0.88} 16.36 & \cellcolor[rgb]{0.89, 0.89, 0.89} 15.76 & \cellcolor[rgb]{0.73, 0.73, 0.73} 28.39 & \cellcolor[rgb]{0.88, 0.88, 0.88} 16.81 & \cellcolor[rgb]{0.90, 0.90, 0.90} 14.75 & \cellcolor[rgb]{0.63, 0.63, 0.63} 33.85 & \cellcolor[rgb]{0.88, 0.88, 0.88} 16.25 & \cellcolor[rgb]{0.88, 0.88, 0.88} 16.29\\ 
  \cmidrule{2-11}
  &  \setGAll   & \cellcolor[rgb]{0.90, 0.90, 0.90} 14.08 & \cellcolor[rgb]{0.94, 0.94, 0.94} 10.14 & \cellcolor[rgb]{0.95, 0.95, 0.95} 9.51 & \cellcolor[rgb]{0.92, 0.92, 0.92} 12.48 & \cellcolor[rgb]{0.92, 0.92, 0.92} 13.01 & \cellcolor[rgb]{0.93, 0.93, 0.93} 11.98 & \cellcolor[rgb]{0.92, 0.92, 0.92} 13.05 & \cellcolor[rgb]{0.94, 0.94, 0.94} 11.05 & \cellcolor[rgb]{0.95, 0.95, 0.95} 9.33\\ 
\bottomrule
\end{tabular}
}
\label{tab:app-test-2}
\end{center}
\end{table*}

On the results using the created training sets, we can answer one question that is not addressed in the main paper: \textbf{Can all the CMs perform well on vocoded training sets?} 
Surprisingly, not. 
By simply switching the training set from {\setTLAshort} to one of the vocoded training set, the {\mA} and {\mB}'s EERs became much worse, and some of them were close to 50\%.  For EERs larger than 50\%, it means that the scores of bona fide trials were lower than those of spoofed trials. 
This is in contrast with the results on {\mC}. EERs on some test sets increased on some training sets, but when training on either {\setTVIVshort}, {\setTVVshort}, or {\setTVVIIshort}, the EERs of {\mC} on test sets in group B were lower than those when training on {\setTLAshort}. 

We hypothesize that {\mA} and {\mB} overfit to the vocoded training set. Their training loss went to zero, but the EERs on other test sets are poor. What leads to the overfitting needs further exploration. According to \citeapp{wang2021investigating}, the conventional CMs picked up the high-frequency spectral information from the ASVspoof 2019 training set, but this does not generalize to other data, leading to higher EERs on other test sets. The SSL-based model seems to avoid this by focusing on the lower half of the frequency band.  Similar spurious information may exist in the vocoded training sets since they are based on the bona fide data of ASVspoof 2019 training set.

\subsection{Additional Results on Contrastive Feature Loss}

Experiment 2 in the main paper showed the results using RawBoost to augment the data. 
Since the goal is to demonstrate the effectiveness of contrastive feature loss, we think the result using RawBoost is sufficient in the page-limited paper.

Here, we report additional results using different configurations for the contrastive feature loss. We use different ways to create the additional view(s):
\begin{itemize}
\item One additional view using compression codec including mp4 and ogg. Mp4 and Ogg were ported through PyDub\footnote{https://github.com/jiaaro/pydub}. We randomly decide whether to use Mp4 or Ogg and which bit rate to use (ranging from 16 to 320 kbps).
\item One additional view using frequency mask. The masking was done at the waveform level using a digital filter, and it is similar to band-pass filtering. The 10th order Butterworth filter was formulated as cascade of second order sections \footnote{Scipy API scipy.signal.sosfiltfilt.}. The bandwidth to mask was randomly decided.
\item Two additional views, one based on RawBoost, the other based on compression codec.
\item Two additional views, one based on RawBoost, the other based on frequency mask.
\item Three additional views using RawBoost, compression, and frequency mask.
\end{itemize}
Similar to the experimental design in the main paper, for each configuration of the contrastive feature loss, we included three systems: one using the vocoded training set {\setTVVshort} and mini-batch with paired bona fide and spoofed trials, one using {\setTVVshort} but randomly composed mini-batch, and one using {\setTLAshort} and randomly composed mini-batch. Note again that when using {\setTLAshort} we do not have pairs of bona fide and spoofed trials.

Results are listed in Table~\ref{tab:app-test-2}. No matter which method is used to create additional view(s), we observe that using the contrastive feature loss and bona fide and spoofed paired mini-batch was helpful in most cases.  
Combination of multi-views using different augmentation methods (e.g., RawBoost and Codec) is promising to achieve better performance on multiple test sets. 

\subsection{Experiment on Impact of Re-sampling}
In Section~\ref{sec:exp1}, we hypothesized that the re-sampling process on the vocoded data may lead to degraded EERs if the CM only see such re-sampled spoofed data. However, the vocoders in {\setTVIshort} and {\setTVIIshort} are not the same, and the evidence in Section~\ref{sec:exp1} is not sufficiently strong.

Hence, we did another experiment, in which the subsets of Voc.v1 and Voc.v2 that contained vocoded data only from HiFiGAN were used.  Note that, although the HiFiGANs were implemented in different toolkits, both follow the official implementation.   
The vocoded utterances from the 24 kHz HiFiGAN were re-sampled to 16 kHz before they were used to train the CM.
The results in Table~\ref{tab:app-sr} still support our tentative suggestion. 
EERs in the right column are in most cases lower than those in the left.

\begin{table}[t!]
\caption{Comparison of vocoded data from HiFiGANs with different sampling rates.}
\centering
\footnotesize
\vspace{-2mm}
\begin{tabular}{crr}
\toprule
 & \shortstack{voc.v1} sub & \shortstack{voc.v2} sub\\ 
EER (\%)  & (HiFiGAN,  & (HiFiGAN, \\ 
(ave. 3 runs) & 24 kHz)  & 16 kHz)  \\ 
\midrule
  LA19eval  & \cellcolor[rgb]{0.83, 0.83, 0.83} 35.42 & \cellcolor[rgb]{0.99, 0.99, 0.99} 18.76\\ 
  LA21eval  & \cellcolor[rgb]{0.81, 0.81, 0.81} 36.81 & \cellcolor[rgb]{0.94, 0.94, 0.94} 25.35\\ 
  DF21eval  & \cellcolor[rgb]{0.88, 0.88, 0.88} 31.34 & \cellcolor[rgb]{0.98, 0.98, 0.98} 20.24\\ 
\midrule
  LA19etrim & \cellcolor[rgb]{0.71, 0.71, 0.71} 43.49 & \cellcolor[rgb]{0.86, 0.86, 0.86} 33.43\\ 
   LA21hid  & \cellcolor[rgb]{0.69, 0.69, 0.69} 44.40 & \cellcolor[rgb]{0.82, 0.82, 0.82} 36.35\\ 
   DF21hid  & \cellcolor[rgb]{0.71, 0.71, 0.71} 43.73 & \cellcolor[rgb]{0.84, 0.84, 0.84} 34.77\\ 
   WaveFake   & \cellcolor[rgb]{0.68, 0.68, 0.68} 45.22 & \cellcolor[rgb]{0.76, 0.76, 0.76} 40.39\\ 
    InWild   & \cellcolor[rgb]{0.66, 0.66, 0.66} 45.99 & \cellcolor[rgb]{0.60, 0.60, 0.60} 49.66\\ 
\midrule
    Pooled     & \cellcolor[rgb]{0.73, 0.73, 0.73} 42.16 & \cellcolor[rgb]{0.81, 0.81, 0.81} 36.99\\ 
\bottomrule
\end{tabular}
\label{tab:app-sr}
\end{table}

\bibliographystyleapp{IEEEbib}
\bibliographyapp{library}

\end{document}